\documentclass[11pt]{article}
\usepackage{amsmath,amssymb}
\usepackage{graphicx}
\usepackage{subcaption}
\usepackage{xcolor}
\usepackage{caption}
\usepackage[T1]{fontenc}
\usepackage{lmodern}
\usepackage{booktabs}
\usepackage{placeins}

\usepackage{url}
\usepackage[hidelinks]{hyperref}
\begin{document}

\setcounter{page}{1}

\pagestyle{plain}

\begin{center}
\Large{\bf Primordial Black Holes and Induced Gravitational Waves from Inflation with History-Dependent Effective Dissipation}\\
\small \vspace{1cm} {Narges
Rashidi}\footnote{n.rashidi@umz.ac.ir } \\
\vspace{0.5cm} $^{}$ Department of Theoretical Physics, Faculty of
Science,
University of Mazandaran,\\
P. O. Box 47416-95447, Babolsar, IRAN\\
\end{center}

\begin{abstract}
We study inflation with an exponential factor multiplying the scalar Lagrangian and depending on the accumulated expansion history through a field-dependent coupling. Variation of the complete action on a finite interval yields an adjoint response with a final-time boundary condition. We derive the background and coupled scalar perturbation equations and examine five benchmarks for an exponential plateau potential and a smooth single-window coupling. The kinetic, direct-coupling and adjoint contributions nearly cancel, producing transient positive minima of the first Hubble slow-roll parameter, $\epsilon_1$, without a comparable reduction in field velocity. The scalar spectra, normalized to the cosmic microwave background (CMB) pivot amplitude, develop localized peaks with amplitudes $0.033$--$0.098$. Under Gaussian statistics and the adopted radiation-era collapse prescription, the the primordial black hole (PBH) abundance peaks span masses from $4.52\times10^{-13}$ to $34.05\,M_\odot$. The lightest-mass benchmark reaches $f_{\rm PBH}\simeq0.99$, allowing nearly all dark matter within the monochromatic approximation. The accompanying scalar-induced gravitational-wave spectra span peak frequencies from $6.78\times10^{-10}$ to $5.92\times10^{-3}\,\mathrm{Hz}$, with selected benchmarks lying within the projected sensitivities of future detectors.

\end{abstract}

\section{Introduction}
\label{sec1}

Cosmic inflation describes an early period of accelerated expansion that addresses the horizon and flatness problems of hot Big Bang cosmology \cite{Guth1981,Linde1982,St80}. Quantum fluctuations generated during this epoch seed primordial curvature perturbations, connecting inflationary dynamics to cosmic microwave background (CMB) anisotropies and the subsequent formation of large-scale structure \cite{Muk88,Sas86}. Measurements of CMB temperature and polarization constrain the amplitude and scale dependence of the scalar spectrum, while searches for primordial $B$-mode polarization constrain the tensor contribution \cite{Planck2018Inflation,BK18}. Planck 2018 observations favor a nearly scale-invariant, red-tilted scalar spectrum, and the BICEP/Keck analysis through the 2018 observing season gives $r_{0.05}<0.036$ at $95\%$ confidence \cite{Planck2018Inflation,BK18}. More recent CMB measurements from the Atacama Cosmology Telescope (ACT), combined with baryon acoustic oscillation measurements from the Dark Energy Spectroscopic Instrument (DESI), provide complementary constraints on inflationary models \cite{ACT,DESI2024BAO}. Their implications for theoretical predictions depend on the adopted dataset combination \cite{ACT,DESI}. These observations establish large-scale benchmarks for inflationary scenarios that generate enhanced primordial fluctuations on smaller scales.

Primordial black holes (PBHs) provide a connection between primordial fluctuations and the dark matter problem. They may form through the gravitational collapse of sufficiently large overdensities in the early Universe, with characteristic masses related to the horizon mass at formation \cite{Hawking1971,CarrHawking1974,Carr1975}. Their possible contribution to dark matter was recognized early on \cite{Chapline1975} and continues to motivate theoretical and observational investigations \cite{Carr2021Constraints,Green2024,Carr2026}. Constraints from evaporation, gravitational lensing, accretion, dynamical effects, and gravitational waves restrict their abundance across a broad range of masses, while an asteroid-mass window remains viable for PBHs to constitute all of the dark matter; the interpretation of these bounds depends on the underlying mass distribution \cite{Carr2021Constraints,Green2024,Carr2026}. Recent studies explore diverse mechanisms for inflationary PBH production. Arya, Jain, and Mishra demonstrated that dissipative dynamics in warm Higgs-G inflation can generate asteroid-mass PBH dark matter together with potentially detectable induced gravitational waves \cite{Arya2024}. Ghosh and Mishra connected asteroid-mass PBH dark matter to induced gravitational-wave signals within the projected reach of future space-based detectors \cite{GhoshMishra2024}. Raatikainen, R\"as\"anen, and Tomberg showed that stochastic fluctuations during ultra-slow-roll inflation can substantially modify the compaction profiles and resulting PBH mass distribution \cite{Raatikainen2024}. More recently, Lorenzoni and collaborators demonstrated that light spectator fields can enhance primordial perturbations through multifield dynamics and facilitate PBH production \cite{Lorenzoni2026}. These developments motivate relating inflationary mechanisms to PBH masses, abundances, and observational signatures.

Appreciable PBH production from primordial overdensities requires substantial scalar-spectrum enhancement on scales much smaller than those probed by the CMB. In scenarios with approximately Gaussian primordial fluctuations and collapse during radiation domination, the required peak amplitude is typically of order $\mathcal{P}_{\zeta}\sim10^{-2}$, several orders of magnitude above the CMB normalization $A_s\simeq2.1\times10^{-9}$, with the precise value depending on the collapse prescription and the shape of the spectrum \cite{Green2024,Byrnes2019}. A nearly scale-invariant, red-tilted extrapolation of the CMB spectrum cannot provide this enhancement, motivating inflationary dynamics that amplify selected small-scale modes while preserving the observed large-scale spectrum. In canonical single-field models, an approximate inflection point or a sufficiently flat region of the potential can produce a transient departure from the slow-roll attractor, including an ultra-slow-roll phase during which curvature perturbations can grow outside the horizon \cite{GermaniProkopec2017,BallesterosTaoso2018}. The amplification depends on this phase and its transitions, requiring the full perturbation dynamics to predict the final spectrum \cite{GermaniProkopec2017,BallesterosTaoso2018,Byrnes2019}. Recent analyses have investigated how smoothing these transitions affects the spectral features and PBH mass distribution \cite{Cole2024}, while potential-reconstruction methods have demonstrated that substantial enhancement can also be achieved within suitably designed slow-roll scenarios \cite{AutieriRedi2025}. These results motivate additional mechanisms for localized enhancement consistent with CMB observations.

Enhanced primordial scalar perturbations also have an accompanying gravitational-wave signature. At second order in cosmological perturbation theory, products of first-order scalar perturbations source tensor modes, generating a stochastic background of scalar-induced gravitational waves as the enhanced modes re-enter the horizon \cite{Ananda2007,Baumann2007,KohriTerada2018}. Consequently, the primordial spectrum responsible for PBH formation also determines an induced gravitational-wave signal whose amplitude and spectral shape depend on the scalar spectrum and the subsequent expansion history \cite{KohriTerada2018,Domenech2021}. This provides a complementary probe of scales inaccessible to CMB anisotropy measurements \cite{Ananda2007,Domenech2021}. For PBH formation during radiation domination, the characteristic PBH mass and the present gravitational-wave frequency are linked through the scale of the enhanced perturbations, with smaller masses generally corresponding to higher frequencies \cite{Domenech2021,Bartolo2019}. Bartolo and collaborators demonstrated that the induced background associated with asteroid-mass PBH dark matter can be tested by LISA \cite{Bartolo2019}, while subsequent work explored this connection across the projected sensitivities of several future detectors \cite{GhoshMishra2024}. At lower frequencies, the NANOGrav 15-year analysis investigated scalar-induced gravitational waves as a possible cosmological interpretation of the nanohertz background \cite{NANO}. Joint PBH and induced-GW predictions thus connect inflationary dynamics with complementary observations.

Effective dissipation offers another way to modify inflationary backgrounds and perturbations. Interactions with additional degrees of freedom can generate effective friction and noise, as demonstrated within the effective field theory of dissipative inflation \cite{LopezNacir2012} and developed further in the open effective field theory of inflation \cite{AguiSalcedo2024}. More generally, integrating out environmental fields can produce nonlocal response terms that retain information about the preceding evolution, motivating the study of history-dependent effective descriptions \cite{Boyanovsky2016}. A complementary phenomenological approach introduces an exponential factor multiplying the scalar-field Lagrangian, with its exponent expressed through a time integral involving the expansion rate and a dissipation function, as proposed by Harko in the context of dissipative quintessence \cite{Harko2023}. Inspired by this structure, we investigate an inflationary model in which the scalar Lagrangian is multiplied by a factor depending on the accumulated expansion history and a field-dependent coupling. The historical functional is varied together with the scalar and gravitational fields, so that its contribution to the background and perturbation equations is retained. Here, effective dissipation refers to this action-level modification of the scalar dynamics; a thermal radiation bath is not assumed. 

We ask whether a smooth, localized coupling in a history-dependent scalar action can produce a small-scale enhancement of the curvature spectrum while retaining viable large-scale inflationary predictions. We formulate the historical dependence on a finite time interval and retain its variation in deriving the background and scalar perturbation equations. This introduces an adjoint contribution whose boundary condition follows from the variational principle. For an exponential plateau potential and a single coupling window, we examine how the kinetic, direct-coupling, and adjoint contributions combine in the first Hubble slow-roll parameter, $\epsilon_1$. In the benchmark solutions studied here, their cancellation produces a transient suppression of this parameter without an equally strong reduction in the field velocity. We then use the resulting scalar spectra to calculate PBH abundances and induced gravitational-wave backgrounds under the stated radiation-era and statistical assumptions. The study thus formulates and numerically explores this historical coupling mechanism and its linked observables.

Our analysis focuses on five benchmark parameter sets, S1--S5, examining their background evolution, primordial spectra, PBH abundances, and induced gravitational-wave signals in comparison with observational constraints and projected detector sensitivities. Section~\ref{sec2} presents the historical action, coupling profile and background equations. Section~\ref{sec3} presents the perturbation framework and the choice of initial state. Section~\ref{sec4} presents the background evolution, primordial spectra and CMB observables, including their parameter dependence. Sections~\ref{sec5} and~\ref{sec6} present the PBH and scalar-induced gravitational-wave results, respectively. Section~\ref{sec7} summarizes the conclusions. Appendix~\ref{secA} provides the quadratic action and canonical boundary data.

\section{Historical coupling and inflationary background}
\label{sec2}

With $M=\kappa^{-1}$ and $\kappa^2=8\pi G$, the action is
\begin{equation}
	S=
	\int d^{4}x\sqrt{-g}
	\left[
	\frac{R}{2\kappa^{2}}
	+
	e^{\Gamma(g_{\mu\nu},\phi,x^\mu)}
	\left(
	-\frac12 g^{\mu\nu}\nabla_\mu\phi\nabla_\nu\phi
	-
	V(\phi)
	\right)
	\right],
	\label{eq1}
\end{equation}
We define its off-background history by the fixed-coordinate ADM integral
\begin{equation}
\Gamma(t,\mathbf x)=\Gamma_i+\int_{t_i}^{t}ds\,
N(s,\mathbf x)K(s,\mathbf x)Q[\phi(s,\mathbf x)].
	\label{eq2}
\end{equation}
Here $N$ is the lapse and $K$ the trace of the extrinsic curvature.
The coordinate label $\mathbf x$ is held fixed; integration along the
ADM normal would instead introduce shift advection and define a different
functional. We vary the action on $[t_i,t_f]$, fixing the homogeneous
initial $\Gamma_i$ and leaving $\Gamma_f$ free.

The homogeneous lapse is retained during variation and set to one
after imposing its constraint. On $ds^2=-dt^2+a^2\delta_{ij}dx^i dx^j$, define
$H=\dot a/a$, $A=e^\Gamma$ and $X=\dot\phi^2/2$. Then
$\dot\Gamma=3HQ$. In uniform-field slicing, $\delta\phi=0$, an
auxiliary representation that retains the full history variation is
\begin{equation}
S_{\rm aux}=S_g+\int dt\,d^3x\,\sqrt h\left\{
e^\Gamma\left(\frac{X}{N}-NV\right)
+\lambda\,[\dot\Gamma-NKQ(\phi)]\right\}.
\label{eq3}
\end{equation}
Here $S_g$ is the ADM gravitational action, specified in
Appendix~\ref{secA}.
Variation with respect to $\Gamma$ gives
\begin{equation}
\dot\lambda+3H\lambda=A(X-V),\qquad
\lambda(t)=-a(t)^{-3}\int_t^{t_f}ds\,a(s)^3A(s)[X(s)-V(s)],
\qquad \lambda_f=0.
\label{eq4}
\end{equation}
The final hypersurface $t_f$ is identified with the end of inflation,
defined by $\epsilon_1(t_f)=1$, and forms part of the finite-interval
definition of the historical action. The condition $\lambda(t_f)=0$
follows from allowing the final value of $\Gamma$ to vary freely.
For the five benchmarks considered here, $N_f-N_*=55$.
The adjoint is not an independent initial datum. The scalar equation is
\begin{equation}
\ddot\phi+3H(1+Q)\dot\phi+V_{,\phi}
+\frac{3H\lambda Q_{,\phi}}{A}=0,
\label{eq5}
\end{equation}
The effective density and pressure, and the Friedmann equations, are
\begin{align}
\rho_\phi&=A(X+V),\qquad
p_\phi=A(1+Q)(X-V)+\lambda Q_{,\phi}\dot\phi,\nonumber\\
3M^2H^2&=\rho_\phi,\qquad
-2M^2\dot H=\rho_\phi+p_\phi.
\label{eq6}
\end{align}
On Eq.~(\ref{eq5}), these quantities satisfy
$\dot\rho_\phi+3H(\rho_\phi+p_\phi)=0$. To specify the background model, we choose the potential and coupling
profile entering these equations. We use an exponential plateau
potential, with the same functional
form as the Starobinsky potential \cite{St80} but a different slope,
\begin{equation}
V(\phi)=V_0\left(1-e^{-\beta\phi/M}\right)^2,\qquad
M=\kappa^{-1}.
\label{eq7}
\end{equation}
Here $\beta$ is dimensionless; we use $\beta=0.024$ for S1--S5. We set $Q_{\rm base}=0$ and use one smooth localized coupling window,
\begin{equation}
\begin{aligned}
Q(\phi)=\frac{Q_{\rm amp}}{4}
&\left[1+\tanh\!\left(\frac{\phi-\phi_L}{w_L}\right)\right]\\
{}\times{}&\left[1-\tanh\!\left(\frac{\phi-\phi_R}{w_R}\right)\right].
\end{aligned}
\label{eq8}
\end{equation}
Here $\phi_L<\phi_R$ locate the left and right transitions and
$w_L,w_R>0$ set their smoothness; all four field scales have the
units of $\phi$, and $Q_{\rm amp}\geq0$ is dimensionless.
We parameterize the transition locations by their center and separation:
\begin{equation}
\begin{aligned}
\phi_c&=\frac{\phi_R+\phi_L}{2}, & \Delta\phi&=\phi_R-\phi_L>0,\\
\phi_L&=\phi_c-\frac{\Delta\phi}{2}, & \phi_R&=\phi_c+\frac{\Delta\phi}{2}.
\end{aligned}
\label{eq9}
\end{equation}
Changing $\phi_c$ translates both transitions together, while
$\Delta\phi$, $w_L$ and $w_R$ specify the common window shape.
The amplitude coefficient $Q_{\rm amp}$ is not generally the
realized maximum. The window has smooth tails rather than compact
support: $Q(\phi)\to0$ as $\phi\to\pm\infty$, but for
$Q_{\rm amp}>0$ it does not vanish at finite field values.
The full history variation of this profile is retained in the
finite-interval action. We label the five single-window parameter sets S1--S5 as listed in
Table~\ref{tab1}. The potential parameters, transition
separation $\Delta\phi$ and smoothness scales $w_L,w_R$ are shared
by all five sets. Only the amplitude $Q_{\rm amp}$ and center
$\phi_c$ vary independently across S1--S5; the transition locations
follow from Eq.~(\ref{eq9}).

\begin{table}[htbp]
\centering\small
\caption{Single-window coupling parameters for S1--S5 in
Eq.~(\ref{eq8}), with $Q_{\rm base}=0$. The separation and both
smoothness scales are common; only $Q_{\rm amp}$ and $\phi_c$ vary
independently. The transition locations $\phi_L,\phi_R$ follow from
Eq.~(\ref{eq9}). $Q_{\rm amp}$ is dimensionless.}
\label{tab1}
\setlength{\tabcolsep}{4pt}
\begin{tabular}{lccccc}
\toprule
Set & $Q_{\rm amp}$ & $\phi_c/M$ & $\Delta\phi/M$ & $w_L/M$ & $w_R/M$\\
\midrule
S1 & $0.0115925$ & $8.49436$ & $1.20926$ & $2.19057$ & $1.45458$\\
S2 & $0.0150533$ & $7.18963$ & $1.20926$ & $2.19057$ & $1.45458$\\
S3 & $0.0176881$ & $6.52554$ & $1.20926$ & $2.19057$ & $1.45458$\\
S4 & $0.0210353$ & $5.92678$ & $1.20926$ & $2.19057$ & $1.45458$\\
S5 & $0.0245589$ & $5.40518$ & $1.20926$ & $2.19057$ & $1.45458$\\
\bottomrule
\end{tabular}

\end{table}

For the background calculation, $N=\ln(a/a_i)$ denotes e-fold time
(the lapse is now one), $y=\phi_N$, $B=\lambda/A$ and
$X=H^2y^2/2$. The Friedmann constraint and exact evolution system are
\begin{equation}
H^2=\frac{A V(\phi)}{3M^2-\tfrac12 A y^2},\qquad
3M^2-\tfrac12 A y^2>0.
\label{eq10}
\end{equation}
\begin{align}
\phi_N&=y,\nonumber\\
y_N&=-[3(1+Q)-\epsilon_1]y
-\frac{V_{,\phi}}{H^2}-\frac{3BQ_{,\phi}}{H},\nonumber\\
\Gamma_N&=3Q(\phi),\nonumber\\
B_N&=\frac{X-V}{H}-3(1+Q)B,\qquad \lambda=AB.
\label{eq11}
\end{align}
Here $\epsilon_1=-H_N/H$ is given below. The boundary condition
$B_f=0$ is equivalently
\begin{equation}
B(N)=e^{-3N-\Gamma(N)}\int_N^{N_f}d\tilde N\,
e^{3\tilde N+\Gamma(\tilde N)}
\frac{V(\tilde N)-X(\tilde N)}{H(\tilde N)},\qquad B(N_f)=0.
\label{eq12}
\end{equation}
The exponential weight can compensate small coupling coefficients in
$BQ_{,\phi}$; small amplitudes alone do not justify a canonical
approximation. In the tails $Q$ and $Q_{,\phi}$ decay and the
source $\Gamma_N=3Q$ becomes small. The accumulated history and
the adjoint are retained and evolved with the same finite-interval
equations; they are not reset outside the window. The slope in Eq.~(\ref{eq7}) is fixed for all five sets. The scalar
power-spectrum amplitude at the pivot scale is normalized to
$A_s=2.1\times10^{-9}$. Reference coupling inputs are given in
Table~\ref{tab1}. The background data and boundary conditions are specified in
Sec.~\ref{sec4}. 
The resulting background equations also identify how the historical
coupling modifies the Hubble evolution. The Hubble hierarchy is
$\epsilon_1=-\dot H/H^2$ and
$\epsilon_2=(\ln|\epsilon_1|)_N$. Equation~(\ref{eq6}) gives
the exact signed decomposition
\begin{align}
\epsilon_1&=\epsilon_K+\epsilon_D+\epsilon_\lambda,\nonumber\\
\epsilon_K&=\frac{AX}{M^2H^2}=\frac{Ay^2}{2M^2},\qquad
\epsilon_D=\frac{AQ(X-V)}{2M^2H^2}
=Q\left(\epsilon_K-\frac32\right),\nonumber\\
\epsilon_\lambda&=\frac{\lambda Q_{,\phi}\dot\phi}{2M^2H^2}
=\frac{ABQ_{,\phi}y}{2M^2H}.
\label{eq13}
\end{align}
For $Q>0$ and $V>X$, the direct contribution $\epsilon_D$ is
negative and can offset $\epsilon_K>0$; the sign of the adjoint
contribution depends on $BQ_{,\phi}y$. A deep positive dip is possible
when $0<\epsilon_K+\epsilon_D+\epsilon_\lambda\ll\epsilon_K$.
This suppresses $\rho_\phi+p_\phi=2M^2H^2\epsilon_1$, while field
motion is measured by $y^2=2M^2\epsilon_K/A$. Consequently a small
$\epsilon_1$ need not require a nearly horizontal $\phi(N)$ or
$V(N)$, since $V_N=V_{,\phi}y$. The corresponding second-parameter contributions are
\begin{equation}
\epsilon_{2a}=\frac{(\epsilon_a)_N}{\epsilon_1},\quad
a=K,D,\lambda,\qquad
\epsilon_2=\sum_a\epsilon_{2a}.
\label{eq14}
\end{equation}
They share the total denominator and are not individual logarithmic
derivatives. Large signed terms can cancel near a small $\epsilon_1$.
The balance is evaluated from the background solution, while the
associated scalar amplification is calculated from the coupled
perturbation equations.

\section{Cosmological perturbations and primordial spectra}
\label{sec3}

We use uniform-field slices, $\delta\phi=0$, on a monotonic segment
$\dot\phi\ne0$, with ADM variables
\begin{equation}
ds^2=-N^2dt^2+h_{ij}(dx^i+N^idt)(dx^j+N^jdt),\qquad
N=1+\alpha,\quad N_i=\partial_i\psi,\quad
h_{ij}=a^2e^{2\zeta}\delta_{ij}.
\label{eq15}
\end{equation}
This slicing and threading are part of the specified continuation.
Writing $\Gamma=\bar\Gamma+g+g_2+O(3)$, where $g$ is linear and
$g_2$ is quadratic in the first-order metric perturbations, gives
\begin{align}
g(t,\mathbf x)&=\int_{t_i}^t ds\,Q
\left(3\dot\zeta-\frac{\partial^2\psi}{a^2}\right),\nonumber\\
g_2(t,\mathbf x)&=\int_{t_i}^t ds\,\frac{Q}{a^2}
\left(2\zeta\partial^2\psi-\partial_i\zeta\partial_i\psi\right).
\label{eq16}
\end{align}
Both vanish initially. Since $\delta Q=0$ on these slices, the history
is fixed by the metric and its initial data rather than introduced
as an independent fundamental field. Let $D=\partial^2\psi/a^2$, $\lambda=\bar\lambda+\ell$ and
$\bar A=e^{\bar\Gamma}$. Variation of the quadratic auxiliary action
in Appendix~\ref{secA} gives, for $k\ne0$,
\begin{align}
&\dot g-Q(3\dot\zeta-D)=0,
\label{eq17}\\
&2M^2(\dot\zeta-H\alpha)+Q\ell=0,
\label{eq18}\\
0={}&\frac{M^2}{2}\left[12H\dot\zeta-4HD
+18H^2\zeta-12H^2\alpha
-\frac{4\partial^2\zeta}{a^2}\right]
+\bar A\left[2X\alpha-(X+V)(3\zeta+g)\right].
\label{eq19}
\end{align}
The Friedmann equation cancels the explicit $\zeta$ terms in the
lapse constraint. On an interval with $Q\ne0$ and $\bar A V\ne0$,
these constraints yield
\begin{align}
D&=3\dot\zeta-\frac{\dot g}{Q},
\label{eq20}\\
\alpha&=\frac{M^2H\dot g/Q-M^2\partial^2\zeta/a^2
-\bar A(X+V)g/2}{\bar A V},
\label{eq21}\\
\ell&=-\frac{2M^2}{Q}(\dot\zeta-H\alpha).
\label{eq22}
\end{align}
The original endpoint data are $g_i=0$ and $\ell_f=0$.
The tanh window has no finite support edges. The undivided
constraints are used numerically, including in the small-$Q$ tails
and in any separately considered zero-coupling limit. Using these constraints, we obtain the coupled scalar dynamics from
the quadratic auxiliary action in Appendix~\ref{secA}.

To simplify the quadratic action, we define
$q=k^2/a^2$, $W=\bar A V$ and $E=\bar A(X+V)$,
together with
\begin{equation}
	\mathcal B=\frac{2M^2H}{Q}\dot g+2M^2q\zeta-Eg.
	\label{eq23}
\end{equation}
Eliminating the constraints on a nonzero-$Q$ interval gives the
real-mode action
\begin{align}
\frac{L^{(2)}_{k,\rm red}}{a^3}={}&
3M^2\dot\zeta^2-\frac{2M^2}{Q}\dot\zeta\dot g
-18M^2H\zeta\dot\zeta
-\frac{27}{2}M^2H^2\zeta^2+M^2q\zeta^2\nonumber\\
&+\bar A(X-V)\left(\frac{9}{2}\zeta^2+3\zeta g+\frac{g^2}{2}\right)
+\bar\lambda\zeta(3\dot g-9Q\dot\zeta)
+\frac{\mathcal B^2}{4W}.
\label{eq24}
\end{align}
For complex modes the products are understood as real Hermitian
combinations. With $\mathcal L=L^{(2)}_{k,\rm red}/a^3$, the equations are
\begin{equation}
\frac{d}{dt}\frac{\partial\mathcal L}{\partial\dot r_a}
+3H\frac{\partial\mathcal L}{\partial\dot r_a}
-\frac{\partial\mathcal L}{\partial r_a}=0,\qquad
r=(\zeta,g)^T.
\label{eq25}
\end{equation}
The kinetic matrix, defined by
$\mathcal L_{\rm kin}=\dot r^T\mathsf K\dot r$, is
\begin{equation}
\mathsf K=\begin{pmatrix}
3M^2&-M^2/Q\\
-M^2/Q&M^4H^2/(Q^2W)
\end{pmatrix},\qquad
\det\mathsf K=\frac{M^4X}{Q^2V}.
\label{eq26}
\end{equation}
The kinetic matrix is positive definite for $X,V>0$ and $Q\ne0$.
The propagation of scalar perturbations is governed by the full coupled
action, including its velocity--gradient mixing terms and the
$M^4q^2\zeta^2/W$ contribution from $\mathcal B^2/(4W)$.
The coupled equations are supplemented by the initial and final
conditions specified below.

For transverse traceless inflationary tensors
$h_{ij}=a^2(e^\gamma)_{ij}$, $\det e^\gamma=1$ leaves the history
source $NK$ unchanged in the pure tensor sector. The action is
\begin{equation}
S_T^{(2)}=\frac{M^2}{8}\int dt\,d^3x\,a^3
\left[\dot\gamma_{ij}\dot\gamma_{ij}
-\frac{\partial_l\gamma_{ij}\partial_l\gamma_{ij}}{a^2}\right].
\label{eq27}
\end{equation}
Each polarization obeys
\begin{equation}
\ddot\gamma_{k}+3H\dot\gamma_{k}+\frac{k^2}{a^2}\gamma_k=0.
\label{eq28}
\end{equation}
For an approximately de Sitter epoch with the tensor adiabatic vacuum,
$\mathcal P_T\simeq2H_k^2/(\pi^2M^2)$ after summing both polarizations.
We evolve the tensor modes on the same background as the scalars.

To specify the scalar initial state consistently with the historical
boundary conditions, we now express the coupled evolution in canonical
variables. The canonical momenta include the factor $a^3$ and use the temporal
surface convention of the auxiliary action. Their undivided expressions
are given in Appendix~\ref{secA}. For one real Fourier
quadrature, let $x=(\zeta,g,p_\zeta,p_g)^T$ and
$x_f=\mathsf U x_i$, where
$\mathsf U^T\mathsf J_4\mathsf U=\mathsf J_4$ and
$\mathsf J_{2n}=\left(\begin{smallmatrix}0&I_n\\-I_n&0\end{smallmatrix}\right)$.
Since $\bar\lambda_f=0$, the history conditions are $g_i=0$ and
$p_{g,f}=0$. For $d_b=\mathsf U_{44}\ne0$, they fix
\begin{equation}
p_{g,i}=-\frac{\mathsf U_{41}\zeta_i+
\mathsf U_{43}p_{\zeta,i}}{d_b},\qquad
x_i=\mathsf E_b R_i,\quad
\mathsf E_b=\begin{pmatrix}
1&0\\0&0\\0&1\\
-\mathsf U_{41}/d_b&-\mathsf U_{43}/d_b
\end{pmatrix},
\label{eq29}
\end{equation}
Here $R_i=(\zeta_i,p_{\zeta,i})^T$. The embedding satisfies
$\mathsf E_b^T\mathsf J_4\mathsf E_b=\mathsf J_2$, so the regular
solution space has one canonical pair $R_i$ per real quadrature.
The history is not assigned an independent vacuum. The classical boundary reduction precedes quantization of this
canonical pair.

On the reduced space, impose
$[\hat R_{i,a},\hat R_{i,b}]=i\hbar(\mathsf J_2)_{ab}$.
A centered Gaussian covariance obeys
\begin{equation}
\mathsf V_i>0,\qquad
\mathsf V_i+\frac{i\hbar}{2}\mathsf J_2\succeq0,
\qquad \det\mathsf V_i\ge\frac{\hbar^2}{4}.
\label{eq30}
\end{equation}
A normalized pure choice \cite{Nicacio2021} is
\begin{equation}
\Psi_i(\zeta_i)=\left(\frac{\alpha_k}{\pi\hbar}\right)^{1/4}
\exp\left[-\frac{\alpha_k+i\beta_k}{2\hbar}\zeta_i^2\right],
\qquad \alpha_k>0,\quad\beta_k\in\mathbb R,
\label{eq31}
\end{equation}
\begin{equation}
\mathsf V_i=\frac{\hbar}{2\alpha_k}
\begin{pmatrix}
1&-\beta_k\\
-\beta_k&\alpha_k^2+\beta_k^2
\end{pmatrix},\qquad \det\mathsf V_i=\frac{\hbar^2}{4}.
\label{eq32}
\end{equation}
We adopt a positive-frequency adiabatic prescription for scalar
perturbations in the early subhorizon regime, specified on the reduced
canonical phase space. The historical momentum is determined by the
final boundary condition, while the initial mode data satisfy the
canonical symplectic normalization. Tensor perturbations are initialized
in the standard Bunch--Davies vacuum. The Gaussian coefficients
$\alpha_k$ and $\beta_k$ are fixed by the scalar initial mode data
and are not additional free parameters. Let $b_\zeta(t)$ be the $\zeta$ row of
$\mathsf E_t=\mathsf U(t,t_i)\mathsf E_b$. The two-point function and
dimensionless power are
\begin{equation}
G_k^+(t,t')=b_\zeta(t)
\left(\mathsf V_i+\frac{i\hbar}{2}\mathsf J_2\right)
 b_\zeta(t')^T.
\label{eq33}
\end{equation}
\begin{equation}
\mathcal P_\zeta(k,t)=\frac{k^3}{2\pi^2}
 b_\zeta(t)\mathsf V_i(k)b_\zeta(t)^T.
\label{eq34}
\end{equation}
These use the Fourier convention
$\langle\zeta_{\mathbf k}\zeta_{\mathbf k'}\rangle
=(2\pi)^3\delta^{(3)}(\mathbf k+\mathbf k')
(2\pi^2/k^3)\mathcal P_\zeta(k)$.
For a pure normalized initial column $\xi_i=(u_i,v_i)^T$,
$u_iv_i^*-u_i^*v_i=i\hbar$, the power reduces to
$k^3|b_\zeta\xi_i|^2/(2\pi^2)$.
At intermediate slices the symplectic form includes both history and
curvature contributions. For a normalized complex solution it gives
\begin{equation}
u_\zeta v_\zeta^*-u_\zeta^*v_\zeta+
u_gv_g^*-u_g^*v_g=i\hbar,
\label{eq35}
\end{equation}
where $v_\zeta,v_g$ are canonical momentum amplitudes. The history
contribution vanishes at the stipulated endpoints; a curvature-only
Wronskian is not conserved in general between them. The adiabatic prescription fixes the normalized scalar state on the
reduced phase space, and the full coupled action determines its
propagation for the specified final slice.
Real temporal surface terms shift canonical momenta and must be
accompanied by the corresponding state phases and transformed endpoint
condition, as recorded in Appendix~\ref{secA}.

\section{Primordial results}
\label{sec4}

We solve Eq.~(\ref{eq11}) in e-fold time with $a_i=1$, $\Gamma_i=0$
and the input velocity $y_i=-M^2V_{,\phi}(\phi_i)/V(\phi_i)$.
The pivot is $k_*=0.05\,\mathrm{Mpc}^{-1}$, with
$p_*=k_*/(a_iH_i)=30$. Its crossing satisfies $k_*=a_*H_*$, and
the ending slice is $N_f=N_*+55$. The background data are chosen
to satisfy the final adjoint condition and the end-of-inflation
condition for each profile S1--S5. Accepted backgrounds have $H^2>0$, a positive denominator in
Eq.~(\ref{eq10}), $\dot\phi\ne0$ on the uniform-field segment and
$0<\epsilon_1<1$ before the end. Scalar and tensor perturbations
are evolved on the same background.
We retain only the mode domain on which the boundary inversion is
regular. The initial slice is selected before
the main window transitions, retaining the full window tails.
The initial-state prescriptions follow Sec.~\ref{sec3}, with the
accumulated history and nonzero window tails retained.

\begin{figure}[htbp]
\centering
\begin{minipage}[t]{0.47\linewidth}
\centering
\includegraphics[width=\linewidth]{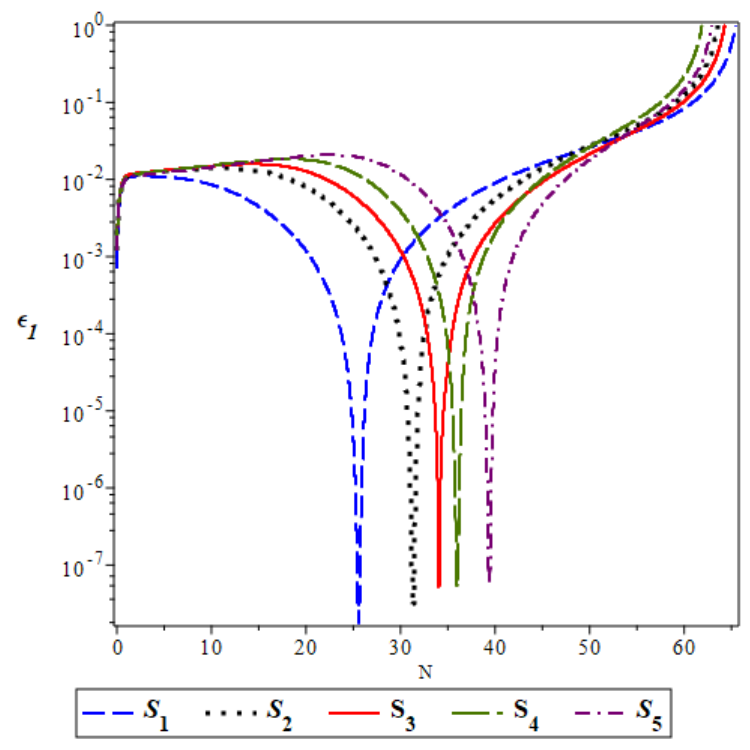}
\end{minipage}\hfill
\begin{minipage}[t]{0.47\linewidth}
\centering
\includegraphics[width=\linewidth]{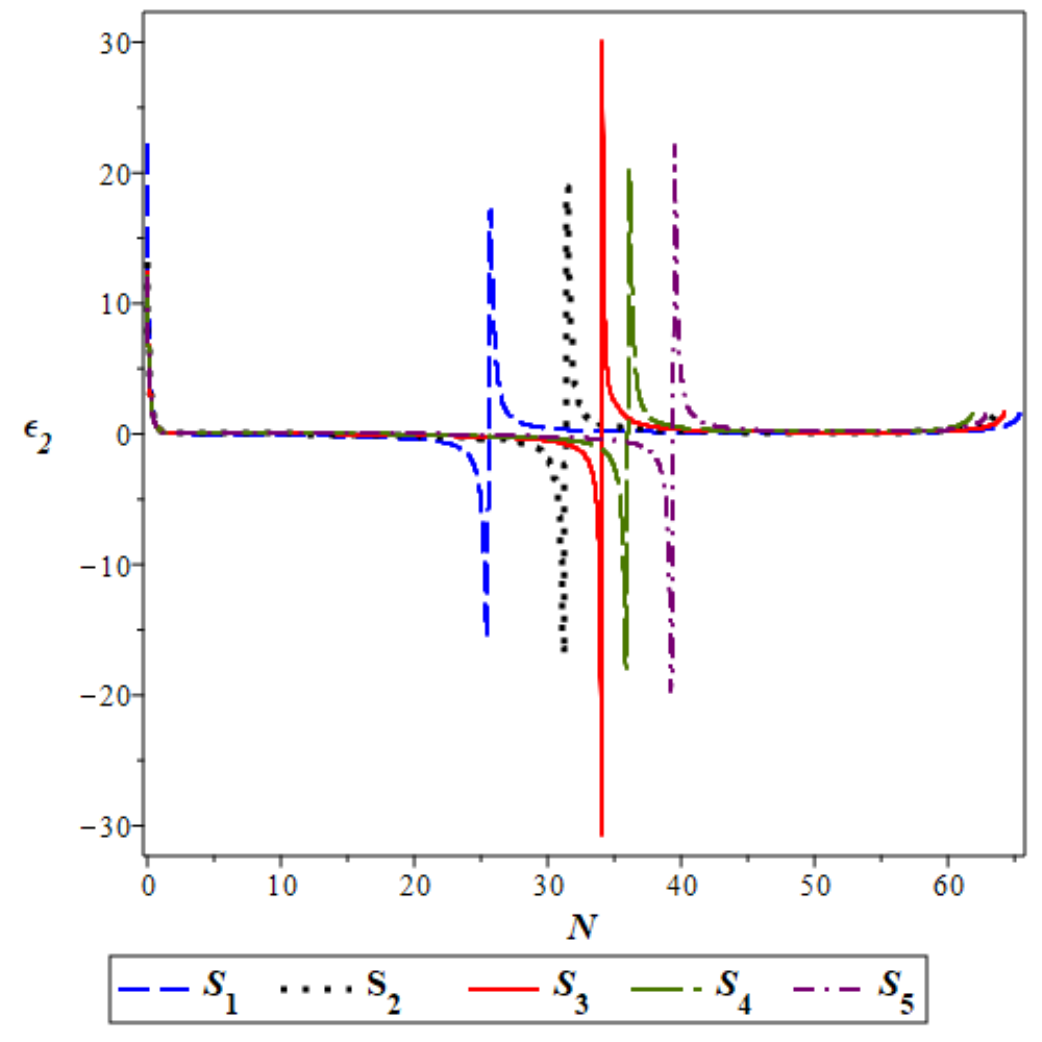}
\end{minipage}

\medskip
\begin{minipage}[t]{0.47\linewidth}
\centering
\includegraphics[width=\linewidth]{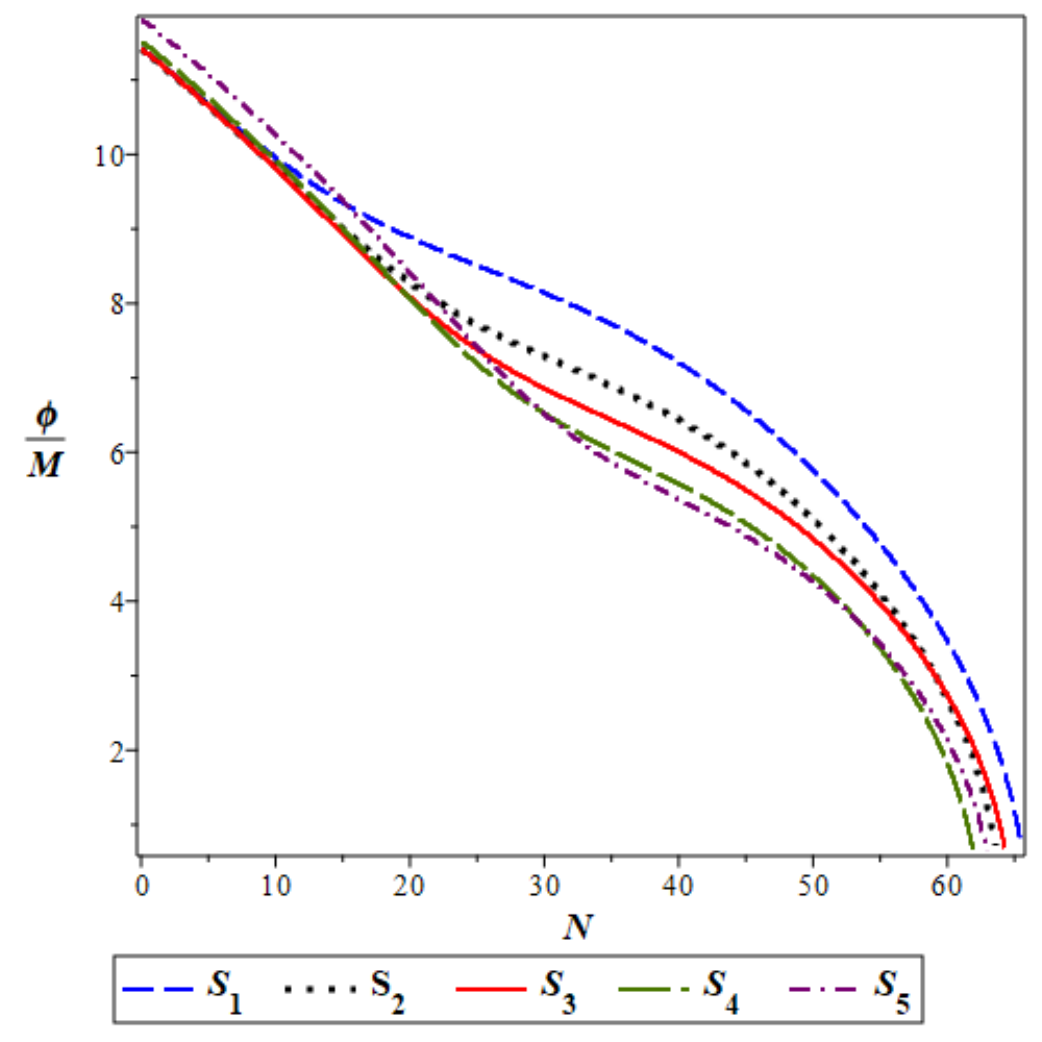}
\end{minipage}\hfill
\begin{minipage}[t]{0.47\linewidth}
\centering
\includegraphics[width=\linewidth]{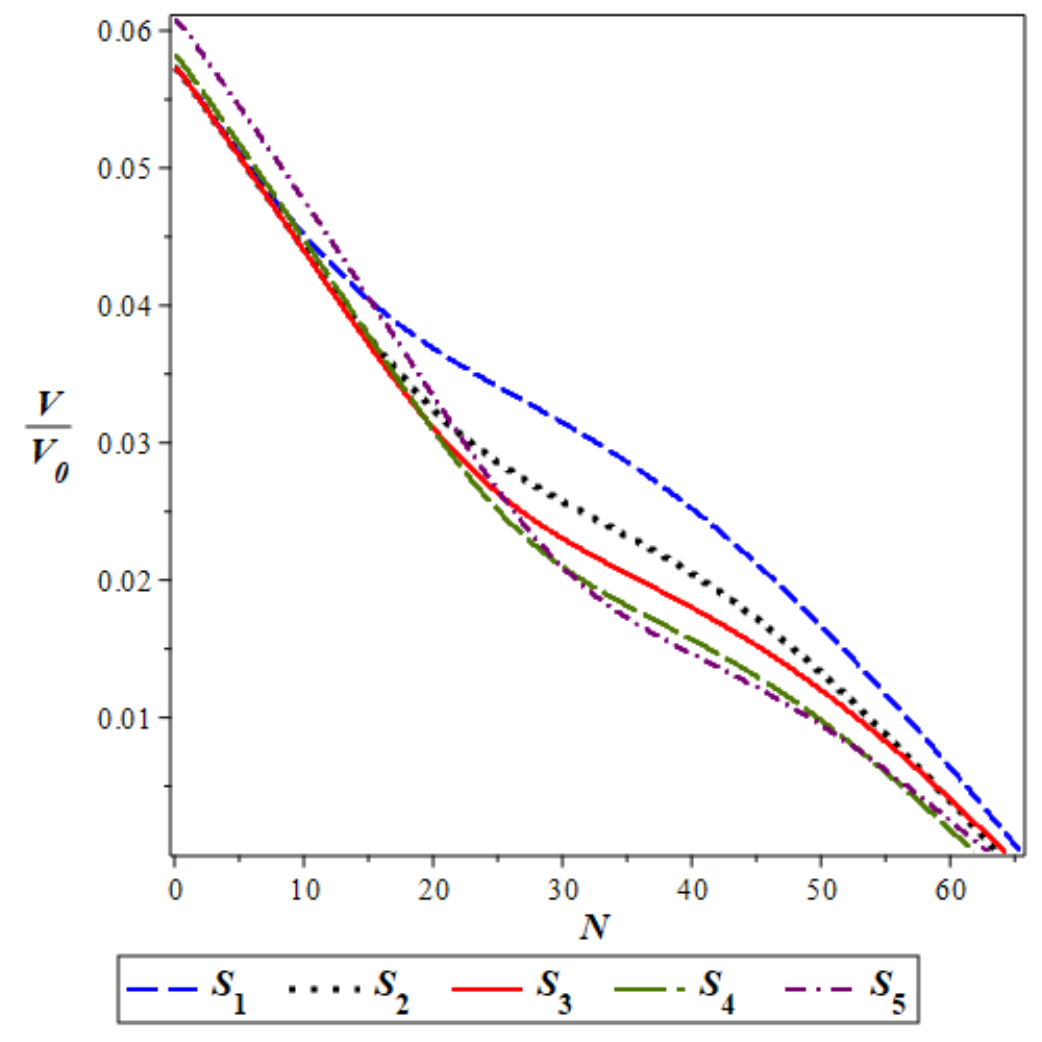}
\end{minipage}
\caption{Background evolution for the benchmark sets S1--S5 in
Table~\ref{tab1}, with the potential in Eq.~(\ref{eq7})
and $\beta=0.024$. The upper left panel shows $\epsilon_1$ on a
logarithmic scale, and the upper right panel shows $\epsilon_2$.
The lower left and lower right panels show $\phi/M$ and $V/V_0$,
respectively. Colors and line styles identify the same benchmark
in every panel.}
\label{fig1}
\end{figure}

For $p=k/(a_iH_i)$, physical normalization in the numerical units
$M=\hbar=a_i=1$ is
\begin{equation}
\mathcal P_\zeta(k)=\frac{H_i^2p^3}{2\pi^2}|Z(p)|^2,\qquad
\mathcal P_T(k)=\frac{H_i^2p^3}{\pi^2}|t(p)|^2,\qquad
\frac{k}{k_*}=\frac{p}{30}.
\label{eq36}
\end{equation}
Here $Z$ and $t$ are the final scalar and per-polarization tensor
amplitudes in the common numerical normalization. Hence
\begin{equation}
n_s=4+2\operatorname{Re}\!\left(\frac{\partial_{\ln p}Z}{Z}\right)_{p_*},
\qquad r=\frac{2|t(p_*)|^2}{|Z(p_*)|^2}.
\label{eq37}
\end{equation}
The background and coupled perturbation equations are solved numerically;
the logarithmic derivative is evaluated from the resulting final modes. With this normalization and mode prescription, we first examine the
background evolution. Figure~\ref{fig1} shows the results for
S1--S5, whose coupling inputs are given in Table~\ref{tab1}.
The five profiles exhibit localized dips in $\epsilon_1$ at different
e-fold times, with rapid variations in $\epsilon_2$, while the field and
its potential decrease monotonically. The reported minima range from
$1.77\times10^{-8}$ to $5.42\times10^{-8}$.

\begin{figure}[p]
\centering
\begin{minipage}[t]{0.45\linewidth}
\vspace{0pt}\centering
\includegraphics[width=\linewidth]{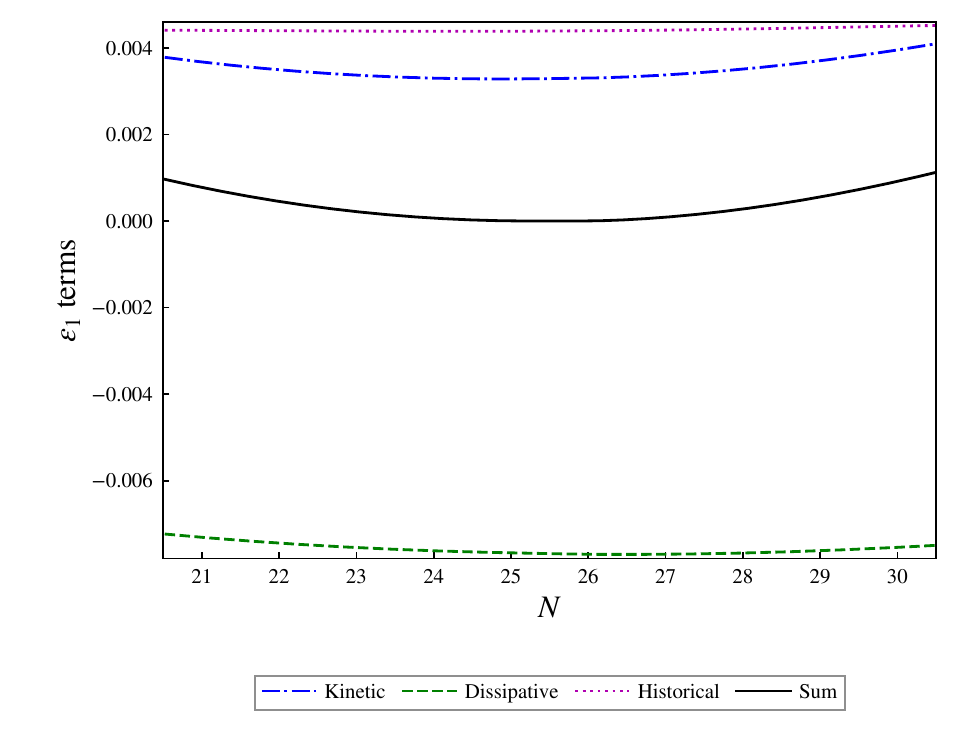}
\end{minipage}\hfill
\begin{minipage}[t]{0.45\linewidth}
\vspace{0pt}\centering
\includegraphics[width=\linewidth]{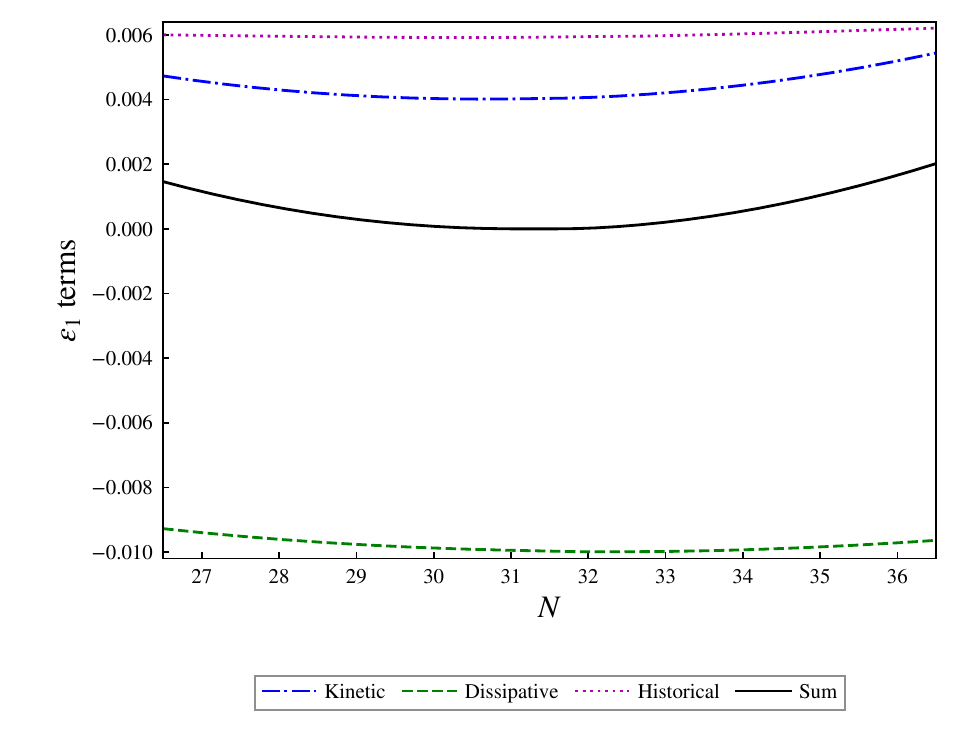}
\end{minipage}

\smallskip
\begin{minipage}[t]{0.45\linewidth}
\vspace{0pt}\centering
\includegraphics[width=\linewidth]{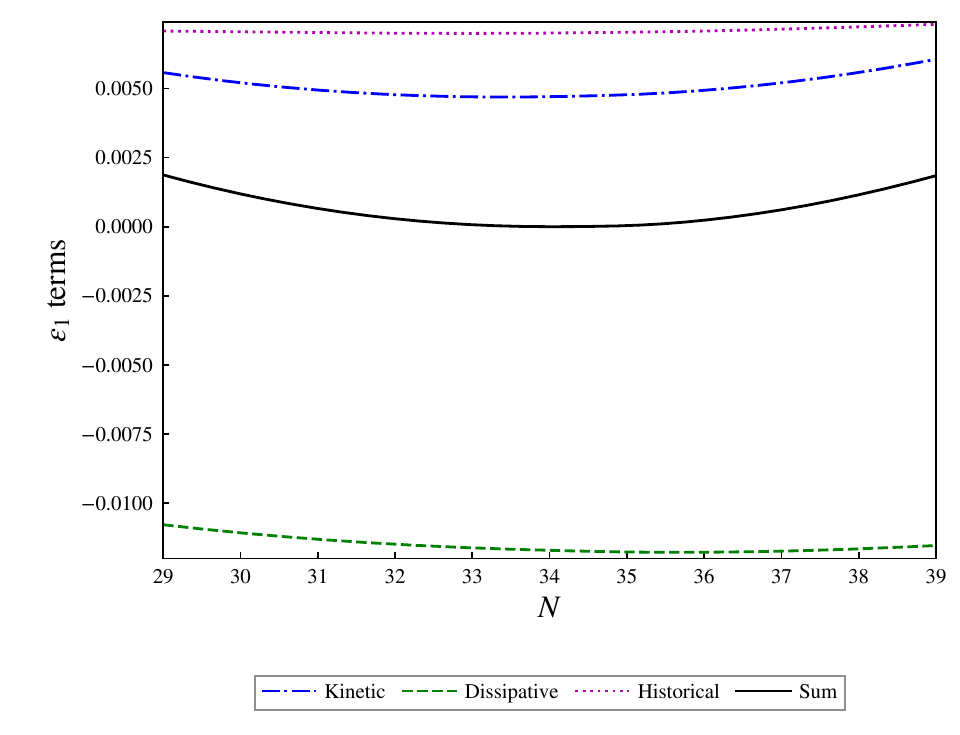}
\end{minipage}\hfill
\begin{minipage}[t]{0.45\linewidth}
\vspace{0pt}\centering
\includegraphics[width=\linewidth]{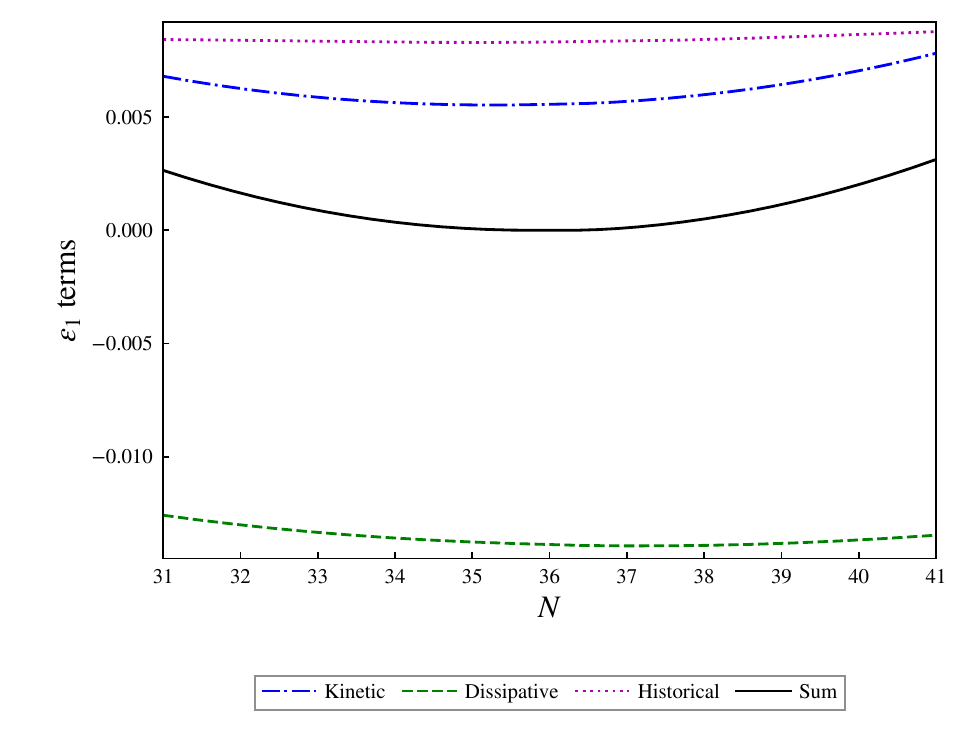}
\end{minipage}

\smallskip
\begin{minipage}[t]{0.45\linewidth}
\vspace{0pt}\centering
\includegraphics[width=\linewidth]{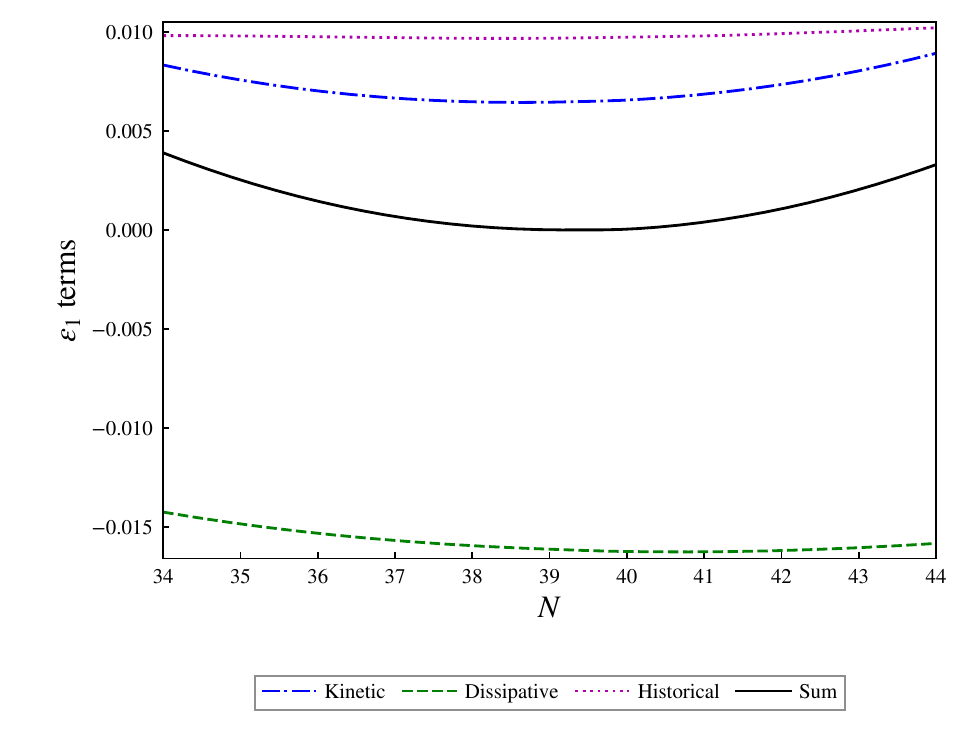}
\end{minipage}
\caption{Signed contributions to $\epsilon_1$ for S1--S5:
S1 and S2 occupy the upper left and upper right, S3 and S4 the middle
left and middle right, and S5 the lower center. The blue dash-dotted,
green dashed and purple dotted curves denote the kinetic,
direct-coupling and historical (adjoint) terms, respectively; the black solid
curve is their sum. Each panel retains its own e-fold interval and
vertical range. Near the dip, the negative direct contribution nearly
cancels the two positive contributions.}
\label{fig2}
\end{figure}

Figure~\ref{fig2} displays the signed
contributions to $\epsilon_1$ for S1--S5. The kinetic and adjoint curves
are positive, whereas the direct-coupling curve is negative. Near the
dip, their individual magnitudes remain of order $10^{-3}$--$10^{-2}$,
while their sum is much smaller. The cancellation occurs progressively
later, from $N_d\simeq25.57$ for S1 to $N_d\simeq39.40$ for S5.
The linear vertical scales make these small positive minima appear
indistinguishable from zero. The nonzero kinetic contribution is
consistent with continued field motion through the dip.
The signed decomposition in Eq.~(\ref{eq13}) provides the
theoretical framework for this balance: the negative direct term can
offset the positive terms while reducing the Hubble evolution rate.
In the finite-interval formulation, the historical term is specifically
$\epsilon_\lambda$, determined by the adjoint equation and its final
condition in Eq.~(\ref{eq4}). Equation~(\ref{eq14}) then
relates the rate of change
of the signed balance to $\epsilon_2$; a small positive total can make
the derivative contributions large. The associated scalar amplification
is obtained from the coupled perturbation calculation presented below.

\begin{figure*}[t]
\centering
\includegraphics[width=1.1\linewidth]{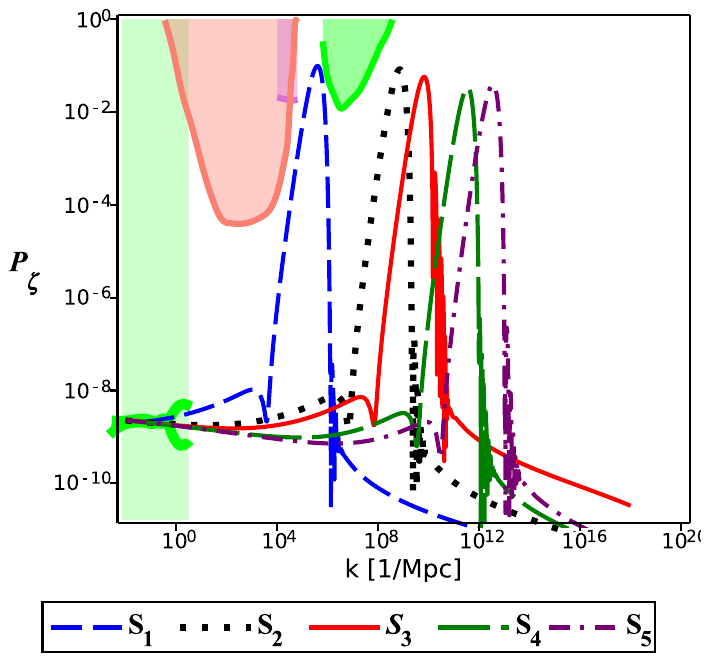}\vspace{-12.5cm}
\caption{Physically normalized scalar power spectra for S1--S5 as
functions of comoving wavenumber $k$. The blue dashed, black dotted,
red solid, olive long-dashed and purple dash-dotted curves denote S1, S2,
S3, S4 and S5, respectively. Their peak locations and amplitudes are
listed in Table~\ref{tab2}. The shaded regions indicate
representative constraints from CMB anisotropies (Planck), CMB
spectral distortions, EPTA observations and BBN
\cite{Planck2018Inflation,Planck2018Parameters,Kohri2014,Chluba2012,Lentati2015,KohriTerada2018}.}
\label{fig3}
\end{figure*}

We next turn to the scalar spectra generated by these backgrounds.
Figure~\ref{fig3} shows the physically normalized
scalar power spectra for S1--S5. Each spectrum exhibits a localized
enhancement above its long-wavelength amplitude. The peak moves from
$k_{\zeta,\rm pk}=3.98153\times10^5\,\mathrm{Mpc}^{-1}$ for S1 to
$3.48433\times10^{12}\,\mathrm{Mpc}^{-1}$ for S5.
Its height decreases from $0.0976935$ for S1 to $0.0331635$ for S4,
then increases to $0.0384728$ for S5. Thus the peak position shifts
monotonically across the five sets, while the peak height does not.
Table~\ref{tab2} lists the scalar, PBH and induced-GW
peak properties. The PBH variance and GW convolution use the full
matched scalar spectrum, including its width and tails, as specified
in Secs.~\ref{sec5} and \ref{sec6}.
The shaded regions provide observational comparisons with CMB
anisotropies~\cite{Planck2018Inflation,Planck2018Parameters},
CMB spectral distortions~\cite{Kohri2014,Chluba2012}, and
constraints associated with EPTA and BBN~\cite{Lentati2015,KohriTerada2018}.
The five benchmark spectra agree with the Planck amplitude band on
CMB scales and exhibit pronounced enhancements at smaller scales,
while remaining outside the exclusion regions associated with CMB
$\mu$-distortions, BBN and EPTA shown in
Fig.~\ref{fig3}.

\begin{table}[htbp]
\centering\small
\caption{Peak properties for S1--S5. Wavenumbers are in $\mathrm{Mpc}^{-1}$ and GW frequencies
are in Hz. Each maximum and its horizontal-axis location are
evaluated separately for the corresponding observable. The PBH
abundance convention is specified in Sec.~\ref{sec5}.}
\label{tab2}
\setlength{\tabcolsep}{3pt}
\begin{tabular}{lcccccc}
\toprule
Set & $k_{\zeta,\rm pk}$ & $\mathcal P_{\zeta,\rm pk}$ &
$M_{\rm PBH,pk}/M_\odot$ & $f_{\rm PBH,pk}$ &
$f_{\rm GW,pk}$ & $(\Omega_{{\rm GW},0}h_{100}^2)_{\rm pk}$\\
\midrule
S1 & $3.98153\times10^{5}$ & $0.0976935$ & $34.0489$ & $7.40010\times10^{-4}$ & $6.77777\times10^{-10}$ & $6.97246\times10^{-8}$\\
S2 & $7.20326\times10^{8}$ & $0.0822519$ & $1.05000\times10^{-5}$ & $0.0260004$ & $1.22468\times10^{-6}$ & $4.91881\times10^{-8}$\\
S3 & $6.71474\times10^{9}$ & $0.0569725$ & $1.26000\times10^{-7}$ & $0.00499995$ & $1.13117\times10^{-5}$ & $2.29313\times10^{-8}$\\
S4 & $3.46675\times10^{11}$ & $0.0331635$ & $4.57000\times10^{-11}$ & $9.01707\times10^{-4}$ & $5.89796\times10^{-4}$ & $7.93707\times10^{-9}$\\
S5 & $3.48433\times10^{12}$ & $0.0384728$ & $4.52400\times10^{-13}$ & $0.990024$ & $5.91852\times10^{-3}$ & $1.06971\times10^{-8}$\\
\bottomrule
\end{tabular}

\end{table}

\begin{table}[htbp]
\centering\small
\caption{CMB observables at $k_*=0.05\,\mathrm{Mpc}^{-1}$ and
$\beta=0.024$ for S1--S5. The scalar amplitude is normalized to
$A_s=2.1\times10^{-9}$ in every set. Peak locations and amplitudes
are listed separately in
Table~\ref{tab2}.}
\label{tab3}
\begin{tabular}{lccccc}
\toprule
Observable & S1 & S2 & S3 & S4 & S5\\
\midrule
$n_s$ & $0.9677$ & $0.9679$ & $0.9678$ & $0.9677$ & $0.9676$\\
$r$ & $0.0210$ & $0.0223$ & $0.0253$ & $0.0254$ & $0.0256$\\
$A_s$ & $2.1\times10^{-9}$ & $2.1\times10^{-9}$ & $2.1\times10^{-9}$ & $2.1\times10^{-9}$ & $2.1\times10^{-9}$\\
\bottomrule
\end{tabular}

\end{table}

\FloatBarrier

The same primordial spectra determine the large-scale CMB observables
and their dependence on the potential slope. At
$k_*=0.05\,\mathrm{Mpc}^{-1}$ we use
$A_s=\mathcal P_\zeta(k_*)$,
$n_s-1=\partial_{\ln k}\ln\mathcal P_\zeta|_*$ and
$r=\mathcal P_T(k_*)/\mathcal P_\zeta(k_*)$, evaluated with the same
state and radiation-matching convention. The scalar amplitude is
normalized to $A_s=2.1\times10^{-9}$ for all five reference sets.
The observables are calculated from the evolved scalar and tensor
spectra. The coupling parameters of S1--S5 are those in
Table~\ref{tab1}. Their reference values at
$\beta=0.024$, also used for the PBH benchmarks, are listed in
Table~\ref{tab3}. The scalar index lies in the narrow interval
$0.9676\le n_s\le0.9679$, while the tensor-to-scalar ratio ranges from
$0.0210$ to $0.0256$. Thus the five sets have closely grouped scalar
tilts but distinguishable tensor amplitudes. 

\begin{figure*}[htbp]
\centering
\includegraphics[width=1.1\linewidth]{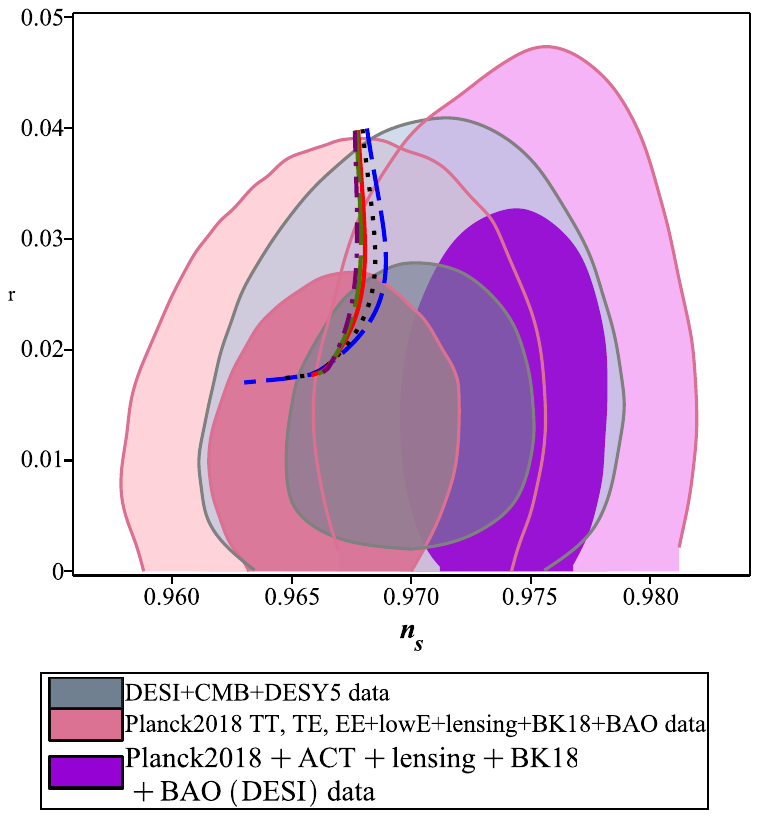}\vspace{-9.5cm}
\caption{Trajectories in the $(n_s,r)$ plane for the S1--S5 coupling
profiles of Table~\ref{tab1}. Curve colors and their
assignment to S1--S5 are the same as in the scalar power spectra
of Fig.~\ref{fig3}: blue, black, red, olive and
purple denote S1, S2, S3, S4 and S5, respectively. Along each trajectory,
$\beta$ varies from $0.020$ to $0.030$, with the coupling parameters
held fixed. The values at $\beta=0.024$ are given in
Table~\ref{tab3}. The shaded regions correspond to the
three data combinations identified in the legend
\cite{Planck2018Inflation,Planck2018Parameters,BK18,ACT,DESI}; inner and
outer contours denote $68\%$ and $95\%$ confidence levels.}
\label{fig4}
\end{figure*}

Figure~\ref{fig4} displays the corresponding trajectories
obtained by varying the potential slope over $0.020\le\beta\le0.030$.
Along each trajectory, the coupling parameters of the corresponding
set are held fixed. This slope scan is distinct from the reference
choice $\beta=0.024$ used for the background, scalar-spectrum and PBH
results. The comparison shows how the CMB predictions respond to the
potential slope while retaining the specified coupling profile. The shaded contours show the DESI+\allowbreak CMB+\allowbreak DESY5, Planck 2018
TT, TE, EE+\allowbreak lowE+\allowbreak lensing+\allowbreak BK18+\allowbreak BAO, and Planck 2018+\allowbreak ACT+\allowbreak lensing+\allowbreak BK18+\allowbreak BAO
(DESI) combinations~\cite{Planck2018Inflation,Planck2018Parameters,BK18,ACT,DESI}.
The inner and outer contours correspond to $68\%$ and $95\%$
confidence levels, respectively. Portions of the model trajectories
overlap the displayed confidence regions, providing a comparison of
the predicted scalar tilt and tensor amplitude with these data
combinations.

\section{Primordial black holes}
\label{sec5}

The post-inflationary calculation assumes Einstein gravity, radiation
domination after reheating and adiabatic transmission of the growing
curvature mode, with negligible decaying modes and no additional
historical source after matching:
\begin{equation}
\zeta_{\rm RD}(\mathbf k)=\zeta_f(\mathbf k),\qquad
\mathcal P_{\zeta}^{\rm RD}(k)=\mathcal P_\zeta(k,t_f).
\label{eq38}
\end{equation}
The accumulated history perturbation $g_f$ may remain nonzero, while
the matching assumes no additional conversion into the growing curvature
mode. The same matched spectrum is used for PBHs and induced waves,
with horizon entry treated in the radiation-dominated approximation. At $aH=R^{-1}$, the radiation-density variance is
\begin{equation}
\sigma^2(R)=\frac{16}{81}\int_0^\infty d\ln q\,
(qR)^4 W^2(qR)\mathcal T^2(qR)\mathcal P_{\zeta}^{\rm RD}(q),
\label{eq39}
\end{equation}
with Gaussian smoothing and the radiation transfer function
\begin{equation}
W(z)=e^{-z^2/2},\qquad
\mathcal T(z)=3\frac{\sin(z/\sqrt3)-(z/\sqrt3)\cos(z/\sqrt3)}
{(z/\sqrt3)^3},\qquad \mathcal T(0)=1.
\label{eq40}
\end{equation}
The collapse threshold depends on the shape of the initial perturbation.
We determine it separately for each benchmark using the analytic
prescription of Musco et al.~\cite{Musco2021Threshold}. Here,
$\delta_c$ denotes the volume-averaged density contrast on superhorizon
scales, extrapolated to horizon crossing during radiation domination.
Starting from the complete numerical curvature spectrum of each
benchmark, we construct the dimensional spectrum
\begin{equation}
P_\zeta(k,\eta)=\frac{2\pi^2}{k^3}\,
\mathcal P_\zeta^{\rm RD}(k)\,T^2(k,\eta),
\label{eq41}
\end{equation}
where the radiation-era transfer function is
\begin{equation}
T(k,\eta)=3\,\frac{\sin u-u\cos u}{u^3},\qquad
u=\frac{k\eta}{\sqrt3},\qquad T(0)=1.
\label{eq42}
\end{equation}
The profile is evaluated outside the horizon at
\begin{equation}
\frac{\hat r_m}{r_H}=10,\qquad r_H=(aH)^{-1}=\eta.
\label{eq43}
\end{equation}
The characteristic radius $\hat r_m$ is determined from the first
maximum of the reconstructed compaction function by solving
\begin{equation}
\int_0^\infty dk\,k^2
\left[(k^2\hat r_m^2-1)\frac{\sin(k\hat r_m)}{k\hat r_m}
+\cos(k\hat r_m)\right]P_\zeta(k,\eta)=0.
\label{eq44}
\end{equation}
The Gaussian shape parameter is then obtained as
\begin{equation}
\alpha_G=-\frac14\left[1+\hat r_m
\frac{\displaystyle\int_0^\infty dk\,k^4\cos(k\hat r_m)P_\zeta(k,\eta)}
{\displaystyle\int_0^\infty dk\,k^3\sin(k\hat r_m)P_\zeta(k,\eta)}\right].
\label{eq45}
\end{equation}
The nonlinearly corrected shape parameter $\alpha$ satisfies
\begin{equation}
F(\alpha)\,[1+F(\alpha)]\,\alpha=2\alpha_G,
\label{eq46}
\end{equation}
with
\begin{equation}
F(\alpha)=\left[1-\frac25e^{-1/\alpha}
\frac{\alpha^{\,1-5/(2\alpha)}}
{\Gamma\!\left(\frac{5}{2\alpha}\right)
-\Gamma\!\left(\frac{5}{2\alpha},\frac1\alpha\right)}\right]^{1/2}.
\label{eq47}
\end{equation}
The corresponding collapse threshold is
\begin{equation}
\delta_c(\alpha)=\frac4{15}e^{-1/\alpha}
\frac{\alpha^{\,1-5/(2\alpha)}}
{\Gamma\!\left(\frac{5}{2\alpha}\right)
-\Gamma\!\left(\frac{5}{2\alpha},\frac1\alpha\right)}.
\label{eq48}
\end{equation}
Here, $\Gamma(s)$ and $\Gamma(s,x)$ denote the complete and upper
incomplete Gamma functions, respectively. The benchmark-dependent
quantities are summarized in Table~\ref{tab4}. The resulting
thresholds are used in the PBH-abundance calculation.

\begin{table}[htbp]
	\centering\small
	\caption{Characteristic compaction radius, shape parameters, and
		collapse threshold for the five benchmarks.}
	\label{tab4}
	\begin{tabular}{lccccc}
		\toprule
		Set & $\hat r_m\,[\mathrm{Mpc}]$ & $k_{\zeta,\rm pk}\hat r_m$ &
		$\alpha_G$ & $\alpha$ & $\delta_c$\\
		\midrule
		S1 & $6.14146\times10^{-6}$  & $2.44524$ & $0.715835$ & $2.23522$  & $0.53544$\\
		S2 & $3.41135\times10^{-9}$  & $2.45728$ & $0.553756$ & $1.56714$  & $0.51427$\\
		S3 & $3.80763\times10^{-10}$ & $2.55672$ & $0.416776$ & $1.08316$  & $0.49345$\\
		S4 & $7.11786\times10^{-12}$ & $2.46759$ & $0.342695$ & $0.849562$ & $0.48082$\\
		S5 & $7.08649\times10^{-13}$ & $2.46917$ & $0.282863$ & $0.674447$ & $0.46977$\\
		\bottomrule
	\end{tabular}
\end{table}

For a Gaussian density PDF $p_R(\delta)$ and constant collapse efficiency
$\gamma_{\rm c}$, we adopt the one-tail probability convention
\begin{equation}
\beta(R)=\gamma_{\rm c}\int_{\delta_c}^\infty p_R(\delta)d\delta
=\frac{\gamma_{\rm c}}{2}\operatorname{erfc}
\left(\frac{\delta_c}{\sqrt2\sigma(R)}\right).
\label{eq49}
\end{equation}
We use the benchmark-dependent thresholds in Table~\ref{tab4} and
$\gamma_{\rm c}=0.2$ within the Gaussian, fixed-efficiency collapse
approximation~\cite{Carr1975,Germani2018}. To express the collapse abundance in terms of PBH mass, we relate
each horizon-entry scale to the mass enclosed within the horizon.
The horizon and PBH masses are
\begin{equation}
M_H(k)=\frac{4\pi}{3}\rho_f H_f^{-3}
=\frac{4\pi M^2}{H_f},\qquad
M_{\rm PBH}(k)=\gamma_{\rm c}M_H(k),\qquad k=a_fH_f.
\label{eq50}
\end{equation}
The subscript $f$ here denotes radiation-era formation, distinct from
the final inflationary slice. Thermal evolution fixes the physical
mass normalization through
\begin{equation}
\rho_f=\frac{\pi^2}{30}g_{\rho,f}T_f^4,\qquad
\frac{a_f}{a_0}=\frac{T_0}{T_f}
\left(\frac{g_{s,0}}{g_{s,f}}\right)^{1/3}.
\label{eq51}
\end{equation}
We use constant radiation-era degrees of freedom
$g_{\rho,f}=g_{s,f}=106.75$. For constant degrees of freedom,
$M_{\rm PBH}\propto k^{-2}$.
For surviving PBHs with negligible accretion and mergers,
\begin{equation}
f_{\rm PBH}(M_{\rm PBH})=
\frac{\rho_{{\rm PBH},0}}{\rho_{{\rm DM},0}}
=\beta(R)\frac{\rho_f}{\rho_{{\rm DM},0}}
\left(\frac{a_f}{a_0}\right)^3.
\label{eq52}
\end{equation}
This is a single-scale abundance estimate. An extended population
instead requires a differential formation mass function before
integrating
$\psi(M_{\rm PBH})=\rho_{{\rm DM},0}^{-1}
d\rho_{{\rm PBH},0}/d\ln M_{\rm PBH}$.
The smoothing and mass mappings use $R=k^{-1}$ at horizon entry.

\begin{figure*}[htp]
\centering
\includegraphics[width=1.1\linewidth]{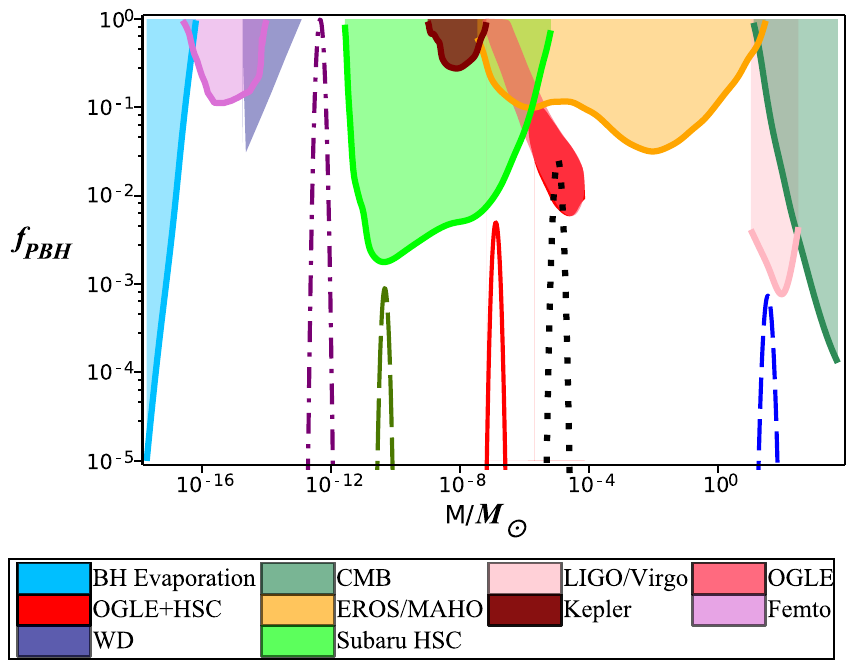}\vspace{-11.5cm}
\caption{Present-day PBH abundance $f_{\rm PBH}$ as a function of
		$M_{\rm PBH}/M_\odot$ for S1--S5. The blue dashed, black dotted,
		red solid, green long-dashed and purple dash-dotted curves denote S1, S2,
		S3, S4 and S5, respectively. The benchmark-dependent collapse
		thresholds are listed in Table~\ref{tab4}, and the peak masses and abundances
		are listed in Table~\ref{tab2}. The red OGLE+HSC patch
		shows an allowed region; the other shaded regions indicate
		representative astrophysical and cosmological upper-limit exclusions,
		labelled in the legend~\cite{Car10,Ali17b,Pou17,Tis07,Nii19,Nii19b,Ali17,Gri13,Carr2021Constraints}.}
	\label{fig5}
\end{figure*}

Figure~\ref{fig5} shows the present-day PBH abundance for
S1--S5, with peak properties listed in Table~\ref{tab2}.
The predicted curves remain outside the exclusion regions displayed
in the figure, while S2 overlaps the allowed OGLE+HSC region.
The observational comparisons include evaporation~\cite{Car10},
CMB accretion constraints~\cite{Ali17b,Pou17}, microlensing searches
\cite{Tis07,Nii19,Nii19b}, LIGO/Virgo~\cite{Ali17} and
Kepler~\cite{Gri13}; an overview is given in
Ref.~\cite{Carr2021Constraints}. The red OGLE+HSC patch denotes
an allowed region, distinct from the shaded upper-limit exclusions.

S1 produces a subdominant population of stellar-mass PBHs, with a
peak near $34.05M_\odot$ and $f_{\rm PBH}\simeq7.4\times10^{-4}$.
Its abundance remains below the displayed LIGO/Virgo bound, making
this benchmark relevant to studies of a possible primordial
contribution to stellar-mass black-hole mergers. 

S2 peaks at $M\simeq1.05\times10^{-5}M_\odot$, with
$f_{\rm PBH}\simeq0.026$. Its overlap with the allowed OGLE+HSC
region indicates compatibility with a PBH interpretation of the
corresponding microlensing events.

S3 and S4 yield subdominant abundances, with respective peaks of
$f_{\rm PBH}\simeq0.005$ at $1.26\times10^{-7}M_\odot$ and
$f_{\rm PBH}\simeq9.02\times10^{-4}$ at
$4.57\times10^{-11}M_\odot$. Both remain below the Subaru HSC
exclusion boundary shown in the figure. Their proximity to this
boundary makes them relevant to improved microlensing constraints.

S5 reaches $f_{\rm PBH}\simeq0.99$ near
$4.52\times10^{-13}M_\odot$, allowing PBHs to account for nearly
all of the dark matter within the monochromatic approximation.
The peak abundances correspond, in this approximation, to dark-matter
fractions of approximately $0.074\%$, $2.6\%$, $0.5\%$, $0.090\%$
and $99\%$ for S1--S5, respectively. These interpretations use
the collapse prescription adopted here and the observational regions
displayed in the figure.

The shift toward smaller peak masses follows the displacement of
the scalar enhancement toward larger wavenumbers, consistent with
$M_{\rm PBH}\propto k^{-2}$. The peak abundances need not follow
the ordering of scalar peak heights: the Gaussian tail depends
exponentially on $\delta_c^2/\sigma^2$, and the thresholds differ
among the benchmarks.

\section{Scalar-induced gravitational waves}
\label{sec6}

We compute the second-order tensor signal sourced by Gaussian scalar
fluctuations after radiation matching. This contribution is separate
from the primordial tensors entering $r$. For modes entering well
after reheating, the late-time radiation-era convolution is
\cite{KohriTerada2018}
\begin{equation}
\Omega_{{\rm GW},{\rm RD}}(k)=\int_0^\infty dv
\int_{|1-v|}^{1+v}du\,
\mathcal K_{\rm RD}(u,v)\mathcal P_{\zeta}^{\rm RD}(ku)\mathcal P_{\zeta}^{\rm RD}(kv).
\label{eq53}
\end{equation}
Here $\Omega_{\rm GW}=\rho_{\rm tot}^{-1}d\rho_{\rm GW}/d\ln k$,
with $u=|\mathbf k-\mathbf q|/k$ and $v=q/k$ for a scalar source
wavevector $\mathbf q$.
With $S=u^2+v^2-3$ and
$G=[4v^2-(1+v^2-u^2)^2]/(4uv)$, its kernel is
\begin{equation}
\begin{aligned}
\mathcal K_{\rm RD}(u,v)={}&\frac{G^2}{12}
\left(\frac{3S}{4u^3v^3}\right)^2
\Biggl\{\left[-4uv+S\ln\left|
\frac{3-(u+v)^2}{3-(u-v)^2}\right|\right]^2\\
&\hspace{40mm}+\pi^2 S^2\Theta(u+v-\sqrt3)\Biggr\}.
\end{aligned}
\label{eq54}
\end{equation}
Here $\Theta$ is the Heaviside function. The kernel includes both
polarizations in the energy-density convention of
Ref.~\cite{KohriTerada2018}; no extra factor of two is applied.
Specifically, for $\delta g_{ij}=a^2\mathfrak h_{ij}/2$ and
orthonormal polarization tensors, the power of one polarization obeys
\begin{equation}
\Omega_{\rm GW}(k,\tau)=\frac1{24}
\left(\frac{k}{\mathcal H}\right)^2
\overline{\mathcal P_{\mathfrak h}(k,\tau)},\qquad
\mathcal H=a'/a,
\label{eq55}
\end{equation}
where the overbar is an oscillation average and the energy fraction
already sums both polarizations. Together with the scalar convolution
and its late-time time-integral average, this fixes the $1/12$
coefficient in Eq.~(\ref{eq54}).
The logarithmic resonance at $u+v=\sqrt3$ is integrable.
Modes entering during reheating require its time-dependent transfer
instead of this analytic radiation limit.

\begin{figure}[htbp]
\centering
\includegraphics[width=1.1\linewidth]{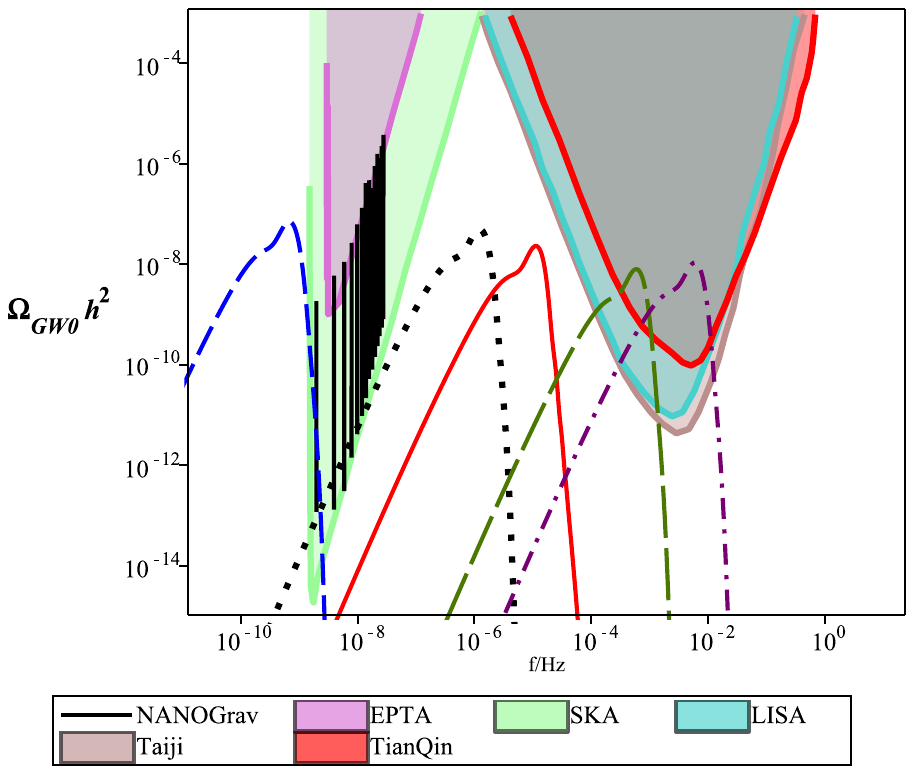}\vspace{-10cm}
\caption{Present-day scalar-induced gravitational-wave spectra
$\Omega_{{\rm GW},0}h_{100}^2$ versus frequency $f$ for S1--S5.
The blue dashed, black dotted, red solid, olive long-dashed and
purple dash-dotted curves denote S1, S2, S3, S4 and S5, respectively.
Peak frequencies and amplitudes are listed in
Table~\ref{tab2}. The comparison curves and shaded regions
refer to NANOGrav~\cite{NANO}, EPTA~\cite{Lentati2015},
SKA~\cite{Moo15}, LISA~\cite{Ama17,Dan97},
Taiji~\cite{Hu17} and TianQin~\cite{Luo16}, as identified in the
legend. Projected detector sensitivities are distinct from existing
observational exclusion limits.}
\label{fig6}
\end{figure}

After production saturates at a radiation-era slice $c$, entropy
conservation gives
\begin{equation}
\Omega_{{\rm GW},0}(k)h_{100}^2=
\Omega_{r,0}h_{100}^2
\frac{g_{\rho,c}}{g_{\rho,0}}
\left(\frac{g_{s,0}}{g_{s,c}}\right)^{4/3}
\Omega_{{\rm GW},{\rm RD}}(k),
\label{eq56}
\end{equation}
where $h_{100}=H_0/(100\,\mathrm{km\,s^{-1}\,Mpc^{-1}})$.
The frequency conversion is
\begin{equation}
f=\frac{ck}{2\pi a_0},\qquad
\frac{f}{\mathrm{Hz}}\simeq1.546\times10^{-15}
\frac{k/a_0}{\mathrm{Mpc}^{-1}}.
\label{eq57}
\end{equation}
We use $a_0=1$ for the reported comoving wavenumbers. The thermal
degrees of freedom are the same as in Sec.~\ref{sec5};
$\Omega_{r,0}h_{100}^2=4.18\times10^{-5}$, with
$g_{\rho,0}=3.36$, $g_{s,0}=3.91$ and
$g_{\rho,c}=g_{s,c}=106.75$.
The PBH and GW calculations use the same matched scalar spectrum.
The GW peak properties are listed in Table~\ref{tab2}.
The peak frequency increases from $6.77777\times10^{-10}\,\mathrm{Hz}$
for S1 to $5.91852\times10^{-3}\,\mathrm{Hz}$ for S5.
The peak values of $\Omega_{{\rm GW},0}h_{100}^2$ range from
$7.93707\times10^{-9}$ to $6.97246\times10^{-8}$.
Figure~\ref{fig6} shows localized enhancements with rising
low-frequency branches and a rapid suppression beyond the main peaks
over the displayed ranges. The peak frequencies move from the
sub-nanohertz range for S1, through the microhertz range for S2 and S3,
to approximately $0.590\,\mathrm{mHz}$ for S4 and
$5.919\,\mathrm{mHz}$ for S5. The peak amplitude decreases from S1
to S4 and then increases for S5, so it does not follow the monotonic
ordering of peak frequencies.
The convolution samples pairs of scalar wavenumbers, so its peak and
width do not follow by relabeling the scalar peak. At fixed shape and
thermal transfer, the Gaussian GW signal scales quadratically with
the scalar power, whose normalization is fixed by
Eq.~(\ref{eq36}). 

The figure compares these spectra with pulsar-timing observations
and detector sensitivities: NANOGrav~\cite{NANO},
EPTA~\cite{Lentati2015}, SKA~\cite{Moo15},
LISA~\cite{Ama17,Dan97}, Taiji~\cite{Hu17} and
TianQin~\cite{Luo16}. S1 and S2 partially overlap the allowed
NANOGrav region and extend into the projected SKA sensitivity region,
making them relevant benchmarks for pulsar-timing observations and
future searches. S3 remains below the sensitivity curves displayed in
Fig.~\ref{fig6}. S4 and S5 overlap the sensitivity regions
of LISA, Taiji and TianQin, offering potential targets for space-based
gravitational-wave observations. None of the five spectra enters the
shaded EPTA region. Together, these results illustrate how the benchmark
profiles produce gravitational-wave signatures across different
frequency bands, with complementary observational prospects.
The detector projections indicate prospective sensitivity and are not
existing exclusion limits.

\section{Conclusions}
\label{sec7}

We have studied inflation with a historical coupling introduced through an exponential factor multiplying the scalar Lagrangian. Retaining the variation of the historical functional on a finite interval yields an adjoint response whose final-time boundary condition follows from the variational principle. We derived the background and coupled scalar perturbation equations and examined five benchmark solutions for an exponential plateau potential and a smooth single-window coupling.

The background solutions exhibit a transient cancellation between the negative direct-coupling term and the positive kinetic and adjoint contributions to the first Hubble slow-roll parameter, $\epsilon_1$. Their individual magnitudes remain of order $10^{-3}$--$10^{-2}$, while the total reaches positive minima of order $10^{-8}$. The field and potential continue to decrease through the dip, showing that this suppression of $\epsilon_1$ does not require a comparable reduction in field velocity. The associated scalar enhancement was calculated from the coupled perturbations. At the reference slope $\beta=0.024$, the benchmarks give $n_s=0.9676$--$0.9679$ and $r=0.0210$--$0.0256$, with $A_s=2.1\times10^{-9}$ at the pivot. Varying $\beta$ from $0.020$ to $0.030$ at fixed coupling parameters produces trajectories that partially overlap the displayed CMB confidence regions. The small-scale scalar peaks have amplitudes $0.0332$--$0.0977$, and the spectra remain outside the displayed exclusion regions associated with CMB spectral distortions, BBN and EPTA.

Using these spectra, the PBH abundance peaks span masses from approximately $34.05\,M_\odot$ for S1 to $4.52\times10^{-13}\,M_\odot$ for S5. S1 yields a subdominant stellar-mass population below the displayed LIGO/Virgo abundance bound. S2 overlaps the allowed OGLE+HSC region, while S3 and S4 remain below the displayed Subaru HSC exclusion boundary. S5 reaches $f_{\rm PBH}\simeq0.99$, allowing nearly all dark matter to consist of PBHs within the monochromatic approximation. These abundance estimates use the adopted Gaussian collapse prescription and benchmark-dependent thresholds; the total fraction of an extended population requires a differential mass function.

The accompanying scalar-induced gravitational-wave spectra peak between $6.78\times10^{-10}$ and $5.92\times10^{-3}\,\mathrm{Hz}$, with peak values of $\Omega_{\rm GW,0}h_{100}^2$ between $7.94\times10^{-9}$ and $6.97\times10^{-8}$. S1 and S2 partially overlap the allowed NANOGrav region and extend into the projected SKA sensitivity region. S4 and S5 overlap the projected sensitivities of LISA, Taiji and TianQin, whereas S3 remains below the displayed sensitivity curves. None enters the shaded EPTA region. These comparisons indicate potential observational relevance across pulsar-timing and space-based frequency bands.

The five benchmarks thus illustrate a historical-coupling mechanism in which a transient balance of background contributions accompanies localized scalar enhancement and its PBH and induced gravitational-wave consequences. The predictions use the specified Gaussian state, adiabatic radiation matching, thermal history and collapse prescription. Within these assumptions, the results connect the inflationary dynamics to complementary comparisons with CMB observations, PBH abundance bounds and gravitational-wave measurements and sensitivities.
\\
\\
\textbf{Acknowledgement}\\
This research was financed by a research grant from the University of Mazandaran.
\\
\\
\textbf{Data Availability Statement} All relevant analytical expressions and
numerical results are included in the manuscript. 
\\
\\
\textbf{Code Availability Statement} No public code repository is associated
with this work. 

\appendix

\section{Quadratic action and canonical boundary data}
\label{secA}

The gravitational action uses the standard ADM temporal convention,
\begin{equation}
	S_g
	=
	\frac{1}{2\kappa^2}
	\int dt\,d^3x\,
	N\sqrt{h}
	\left(
	{}^{(3)}R
	+
	K_{ij}K^{ij}
	-
	K^2
	\right),
	\label{eq58}
\end{equation}
Our extrinsic-curvature sign convention is
\begin{equation}
	K_{ij}
	=
	\frac{1}{2N}
	\left(
	\dot h_{ij}
	-
	D_iN_j
	-
	D_jN_i
	\right),
	\qquad
	K=h^{ij}K_{ij}.
	\label{eq59}
\end{equation}
For the metric in Eq.~(\ref{eq15}), define
\begin{equation}
d_2=\frac{\partial_i\zeta\,\partial_i\psi-2\zeta\partial^2\psi}{a^2},
\qquad
E_2=\frac{(\partial_i\partial_j\psi)^2-(\partial^2\psi)^2}{a^4}.
\label{eq60}
\end{equation}
Up to spatial boundary terms, the quadratic auxiliary action is
\begin{equation}
S^{(2)}=\int dt\,d^3x\,a^3
\left(\mathcal L_g^{(2)}+\mathcal L_m^{(2)}
+\mathcal L_\lambda^{(2)}\right),
\label{eq61}
\end{equation}

\begin{align}
\mathcal L_g^{(2)}=\frac{M^2}{2}\Bigl\{
&-6\dot\zeta^2+4\dot\zeta D+4H d_2+E_2
+(3\zeta-\alpha)(-12H\dot\zeta+4HD)\nonumber\\
&-6H^2\left(\frac{9}{2}\zeta^2-3\alpha\zeta+\alpha^2\right)
+\frac{1}{a^2}\left[-4(\zeta+\alpha)\partial^2\zeta
-2(\partial_i\zeta)^2\right]\Bigr\},
\label{eq62}\\
\mathcal L_m^{(2)}={}&\bar A\left\{X\alpha^2
-(X+V)\alpha(3\zeta+g)
+(X-V)\left(\frac{9}{2}\zeta^2+3\zeta g+\frac{g^2}{2}\right)\right\},
\label{eq63}\\
\mathcal L_\lambda^{(2)}={}&
\ell\,[\dot g-Q(3\dot\zeta-D)]\nonumber\\
&+\bar\lambda\left\{-Q\left(2\zeta D
-\frac{\partial_i\zeta\partial_i\psi}{a^2}\right)
+3\zeta[\dot g-Q(3\dot\zeta-D)]\right\}.
\label{eq64}
\end{align}
Terms linear in independent second-order fields multiply background
equations. For the history, their temporal boundary contribution
vanishes because $g_{2i}=0$ and $\bar\lambda_f=0$.
Thus this expression includes the quadratic history variation without
an additional independent $g_2$ in the reduced action. The canonical momenta, including $a^3$, admit the undivided forms
\begin{equation}
\begin{aligned}
p_\zeta&=a^3\left(2M^2D-18M^2H\zeta
-9\bar\lambda Q\zeta\right),\\
p_g&=a^3(\ell+3\bar\lambda\zeta).
\end{aligned}
\label{eq65}
\end{equation}
The constraints give the regular velocities for $X>0$,
\begin{equation}
\begin{gathered}
D=\frac{p_\zeta/a^3+18M^2H\zeta+9\bar\lambda Q\zeta}{2M^2},
\qquad \ell=p_g/a^3-3\bar\lambda\zeta,\\
\dot\zeta=\frac{M^2H^2D-M^2Hq\zeta+HEg/2
+WQ\ell/(2M^2)}{\bar A X},\qquad
\dot g=Q(3\dot\zeta-D).
\end{gathered}
\label{eq66}
\end{equation}
These expressions contain no $1/Q$ and remain regular through the
smooth transitions. The tail values of $Q$ are retained in the
history evolution. In an exactly zero-coupling limit,
$\dot g=0$ while its conjugate response may evolve. On a nonzero-$Q$ interval write
$L_k=a^3(\dot r^T\mathsf K\dot r+\dot r^T\mathsf C r+
r^T\mathsf F r)$. With $L_b=\bar A(X-V)$, the coefficients extracted
from Eq.~(\ref{eq24}) are
\begin{align}
\mathsf C&=\begin{pmatrix}
-18M^2H-9\bar\lambda Q&0\\
2M^4Hq/(QW)+3\bar\lambda&-M^2HE/(QW)
\end{pmatrix},\nonumber\\
\mathsf F_{11}&=-\frac{27}{2}M^2H^2+M^2q+
\frac92 L_b+\frac{M^4q^2}{W},\nonumber\\
\mathsf F_{12}&=\mathsf F_{21}=\frac12
\left(3L_b-\frac{M^2qE}{W}\right),\qquad
\mathsf F_{22}=\frac12L_b+\frac{E^2}{4W}.
\label{eq67}
\end{align}
With $e=\epsilon_K$ and $W=M^2H^2(3-e)$,
\begin{equation}
\mathsf K^{-1}=\frac{1}{M^2e}
\begin{pmatrix}
1&(3-e)Q\\
(3-e)Q&3(3-e)Q^2
\end{pmatrix}.
\label{eq68}
\end{equation}
The momentum and Hamiltonian are
$p=a^3(2\mathsf K\dot r+\mathsf C r)$ and
\begin{equation}
\mathcal H_k=\frac{1}{4a^3}
(p-a^3\mathsf C r)^T\mathsf K^{-1}(p-a^3\mathsf C r)
-a^3r^T\mathsf F r.
\label{eq69}
\end{equation}
This Legendre transform and the undivided momenta generate the same
regular first-order evolution. For the boundary embedding in
Eq.~(\ref{eq29}), let $\mathsf L$ select
$(\zeta,p_\zeta)$ and $\mathsf B_f=\mathsf L\mathsf U\mathsf E_b$.
At the final slice $\delta p_{g,f}=0$, so symplectic conservation gives
$\mathsf B_f^T\mathsf J_2\mathsf B_f=\mathsf J_2$.
This need not hold for the curvature-only projection at an intermediate
slice. For $\mathsf B_f=\left(\begin{smallmatrix}a_b&b_b\\c_b&d_b'\end{smallmatrix}\right)$
and $b_b\ne0$, the normalized reduced kernel is
\begin{equation}
\mathcal K_b(\zeta_f,\zeta_i)=
\frac{e^{i\vartheta_b}}{\sqrt{2\pi\hbar|b_b|}}
\exp\left[\frac{i}{2\hbar b_b}
\left(d_b'\zeta_f^2+a_b\zeta_i^2-2\zeta_f\zeta_i\right)\right].
\label{eq70}
\end{equation}
The phase specifies the continuous metaplectic branch; $b_b=0$
requires another generating chart. This quantizes the reduced
symplectic map rather than projecting onto sharp auxiliary boundary
eigenstates. A real temporal surface functional $F_{\rm surf}$ shifts
$p\mapsto p+\partial_rF_{\rm surf}$ and the kernel by
$e^{iF_{{\rm surf},f}/\hbar}\mathcal K_b
e^{-iF_{{\rm surf},i}/\hbar}$.
The transformed endpoint condition is
$p'_{g,f}-\partial_gF_{{\rm surf},f}=0$.

\end{document}